\documentclass[aps,prl,preprint,nopacs,superscriptaddress]{revtex4}
\usepackage{graphicx}
\usepackage{verbatim}
\usepackage{relsize}
\usepackage{mathrsfs}
\usepackage{color}
\usepackage{ulem}
\usepackage{amsmath,amsfonts,amssymb}
\usepackage{graphicx}

\usepackage{bbding}
\usepackage{pifont}
\usepackage{wasysym}
\usepackage{amssymb}

\def \beq {\begin{equation}}
\def \eeq {\end{equation}}
\begin{document}

\title{An antiferromagnetic diode effect in even-layered MnBi$_2$Te$_4$}

\author{Anyuan Gao}\affiliation{\footnotesize Department of Chemistry and Chemical Biology, Harvard University, Massachusetts 02138, USA}

\author{Shao-Wen Chen}\affiliation{\footnotesize Department of Physics, Harvard University, Cambridge, MA 02138, USA}

\author{Barun Ghosh}\affiliation{\footnotesize Department of Physics, Northeastern University, Boston, MA 02115, USA}

\author{Jian-Xiang Qiu}\affiliation{\footnotesize Department of Chemistry and Chemical Biology, Harvard University, Massachusetts 02138, USA}

\author{Yu-Fei Liu}\affiliation{\footnotesize Department of Chemistry and Chemical Biology, Harvard University, Massachusetts 02138, USA}\affiliation{\footnotesize Department of Physics, Harvard University, Cambridge, MA 02138, USA}

\author{Yugo Onishi}\affiliation{\footnotesize Department of Physics, Massachusetts Institute of Technology, Cambridge, MA 02139, USA}

\author{Chaowei Hu}\affiliation{\footnotesize Department of Physics and Astronomy and California NanoSystems Institute, University of California, Los Angeles, Los Angeles, CA 90095, USA.}

\author{Tiema Qian}\affiliation{\footnotesize Department of Physics and Astronomy and California NanoSystems Institute, University of California, Los Angeles, Los Angeles, CA 90095, USA.}

\author{Damien B\'erub\'e}\affiliation{\footnotesize Department of Chemistry and Chemical Biology, Harvard University, Massachusetts 02138, USA}

\author{Thao Dinh}\affiliation{\footnotesize Department of Chemistry and Chemical Biology, Harvard University, Massachusetts 02138, USA}\affiliation{\footnotesize Department of Physics, Harvard University, Cambridge, MA 02138, USA}

\author{Houchen Li}\affiliation{\footnotesize Department of Chemistry and Chemical Biology, Harvard University, Massachusetts 02138, USA}

\author{Christian Tzschaschel}\affiliation{\footnotesize Department of Chemistry and Chemical Biology, Harvard University, Massachusetts 02138, USA}

\author{Seunghyun Park}\affiliation{\footnotesize Department of Physics, Harvard University, Cambridge, MA 02138, USA}

\author{Tianye Huang}\affiliation{\footnotesize Department of Chemistry and Chemical Biology, Harvard University, Massachusetts 02138, USA}

\author{Shang-Wei Lien}\affiliation{\footnotesize Department of Physics, National Cheng Kung University, Tainan 70101, Taiwan}\affiliation{\footnotesize Center for Quantum Frontiers of Research and Technology (QFort), Tainan 70101, Taiwan}\affiliation{\footnotesize Physics Division, National Center for Theoretical Sciences, Taipei 10617, Taiwan}

\author{Zhe Sun}\affiliation{\footnotesize Department of Chemistry and Chemical Biology, Harvard University, Massachusetts 02138, USA}\affiliation{\footnotesize Department of Physics, Boston College, Chestnut Hill, MA, USA}

\author{Sheng-Chin Ho}\affiliation{\footnotesize Department of Chemistry and Chemical Biology, Harvard University, Massachusetts 02138, USA}

\author{Bahadur Singh}\affiliation{\footnotesize Department of Condensed Matter Physics and Materials Science,
Tata Institute of Fundamental Research, Colaba, Mumbai 400005, India}

\author{Kenji Watanabe}\affiliation{\footnotesize Research Center for Functional Materials, National Institute for Materials Science, 1-1 Namiki, Tsukuba 305-0044, Japan}

\author{Takashi Taniguchi}\affiliation{\footnotesize International Center for Materials Nanoarchitectonics, National Institute for Materials Science,  1-1 Namiki, Tsukuba 305-0044, Japan}

\author{David C. Bell}\affiliation{\footnotesize Harvard John A. Paulson School of Engineering and Applied Sciences, Harvard University, Cambridge, Massachusetts 02138, USA}\affiliation{\footnotesize Center for Nanoscale Systems, Harvard University,Cambridge, Massachusetts 02138, USA}

\author{Arun Bansil}
\affiliation{\footnotesize Department of Physics, Northeastern University, Boston, MA 02115, USA}\affiliation{\footnotesize Quantum Materials and Sensing Institute, Northeastern University, Burlington, MA 01803, USA}
\author{Hsin Lin}
\affiliation{\footnotesize Institute of Physics, Academia Sinica, Taipei 11529, Taiwan}

\author{Tay-Rong Chang}\affiliation{\footnotesize Department of Physics, National Cheng Kung University, Tainan 70101, Taiwan}\affiliation{\footnotesize Center for Quantum Frontiers of Research and Technology (QFort), Tainan 70101, Taiwan}\affiliation{\footnotesize Physics Division, National Center for Theoretical Sciences, Taipei 10617, Taiwan}

\author{Amir Yacoby}\affiliation{\footnotesize Department of Physics, Harvard University, Cambridge, MA 02138, USA}

\author{Ni Ni}\affiliation{\footnotesize Department of Physics and Astronomy and California NanoSystems Institute, University of California, Los Angeles, Los Angeles, CA 90095, USA.}

\author{Liang Fu}\affiliation{\footnotesize Department of Physics, Massachusetts Institute of Technology, Cambridge, MA 02139, USA}

\author{Qiong Ma}\affiliation{\footnotesize Department of Physics, Boston College, Chestnut Hill, MA, USA}

\author{Su-Yang Xu\footnote{Corresponding author: suyangxu@fas.harvard.edu}}\affiliation{\footnotesize Department of Chemistry and Chemical Biology, Harvard University, Massachusetts 02138, USA}

\date{\today}
\maketitle

\textbf{In a PN junction, the separation between positive and negative charges leads to diode transport. In the past few years, the intrinsic diode transport in noncentrosymmetric polar conductors has attracted great interest, because it suggests novel nonlinear applications and provides a symmetry-sensitive probe of Fermi surface. Recently, such studies have been extended to noncentrosymmetric superconductors, realizing the superconducting diode effect. Here, we show that, even in a centrosymmetric crystal without directional charge separation, the spins of an antiferromagnet (AFM) can generate a spatial directionality, leading to an AFM diode effect. We observe large second-harmonic transport in a nonlinear electronic device enabled by the compensated AFM state of even-layered MnBi$_2$Te$_4$. We also report a novel electrical sum-frequency generation (SFG), which has been rarely explored in contrast to the well-known optical SFG in wide-gap insulators. We demonstrate that the AFM enables an in-plane field-effect transistor and harvesting of wireless electromagnetic energy. The electrical SFG establishes a powerful method to study nonlinear electronics built by quantum materials. The AFM diode effect paves the way for potential device concepts including AFM logic circuits, self-powered AFM spintronics and other applications that potentially bridge nonlinear electronics with AFM spintronics.}

\vspace{0.5cm}
\textbf{Introduction}

Antiferromagnets (AFMs) are elementary yet fascinating magnetic materials. They are internally magnetic, but the zero net magnetization makes the magnetism externally invisible. Although the internal magnetic structure does not manifest as a global magnetization, it can profoundly affect many other macroscopic properties, leading to novel physics: In strongly correlated systems, the anti-parallel spin structure promotes virtual hopping, making AFM a favorable ground state in un-doped Mott insulators \cite{lee2006doping}. In topological physics, the internal AFM spin structures can lead to novel topological phases such as the AFM topological insulator \cite{mong2010antiferromagnetic}, AFM Dirac semimetals \cite{tang2016dirac}, and the Axion insulator \cite{qi2008topological,essin2009magnetoelectric}. The absence of net magnetization and the fast dynamics make them the ideal candidates for next-generation spintronic platforms, which motivates the fast-developing AFM spintronics \cite{jungwirth2016antiferromagnetic}. Therefore, discovering novel transport phenomena in AFM materials is of great interest. 

The intrinsic second-order transport have been discovered in noncentrosymmetric conductors \cite{rikken2001electrical, rikken2005magnetoelectric, tokura2018nonreciprocal,ma2021topology,orenstein2021topology,moore2010confinement,gao2014field,sodemann2015quantum,rectification,oh2016field, yasuda2016large,he2018bilinear,godinho2018electrically, ma2019observation,Kang2019nonlinear,kumar2021room,sinha2022berry, isobe2020high,yasuda2020large, zhao2020magnetic, tsirkin2022separation, Lai2021third,he2022graphene,itahashi2022giant, zhang2022non,ando2015topological,wu2022the, gao2023quantum, wang2023quantum, zhang2020higher, Wang2021Intrinsic,liu2021intrinsic, holder2021mixed, lahiri2022intrinsic, Smith2022momentum, ma2022anomalous, zhang2022efficient, Zhang2022diodic, hayami2022nonlinear, kaplan2022unification, huang2023nonlinear, huang2023Scaling, atencia2023disorder}. Such intrinsic second-order transport leads to a diode-like $I-V$ character (e.g. if $V=aI+bI^2$, then $V(+I)\neq -V(-I)$). In particular, noncentrosymmatric (including ferromagnetic) materials show magneto-chiral anisotropy, nonreciprocal magnetoresistance, or unidirectional magnetoresistance induced by the nonlinear Drude effect \cite{rectification,yasuda2016large,he2018bilinear}; They show the nonlinear Hall effect without magnetic field due the Berry curvature dipole \cite{moore2010confinement,sodemann2015quantum, ma2019observation,Kang2019nonlinear,kumar2021room}; They also support large second-order transport without magnetic field due to skew scattering \cite{isobe2020high, he2022graphene,itahashi2022giant}. More recently, such studies have been extended to noncentrosymmetric polar superconductors, realizing the superconducting diode effect \cite{ando2015topological,wu2022the}. Fundamentally, in these systems, the spatial directionality required for diode transport is provided by the noncentrosymmetric charge distribution. 

Distinct from the noncentrosymmetric materials, recently, the nonlinear transport in time-reversal breaking (magnetic) systems has attracted great interest. This emerging direction has exciting theoretical possibilities, ranging from the quantum metric effects to the possibility of nonlinear effect in magnetic insulators, disorder scattering, etc \cite{gao2014field, zhang2020higher, Wang2021Intrinsic,liu2021intrinsic, holder2021mixed, lahiri2022intrinsic, Smith2022momentum, ma2022anomalous, hayami2022nonlinear, kaplan2022unification, kaplan2023general, huang2023nonlinear, huang2023Scaling, atencia2023disorder}. The quantum metric effect has been recently studied experimentally in \cite{gao2023quantum, wang2023quantum}. However, there are many more open questions, awaiting for both new theoretical and experimental efforts.

\begin{figure*}[t]
\includegraphics[width=14cm]{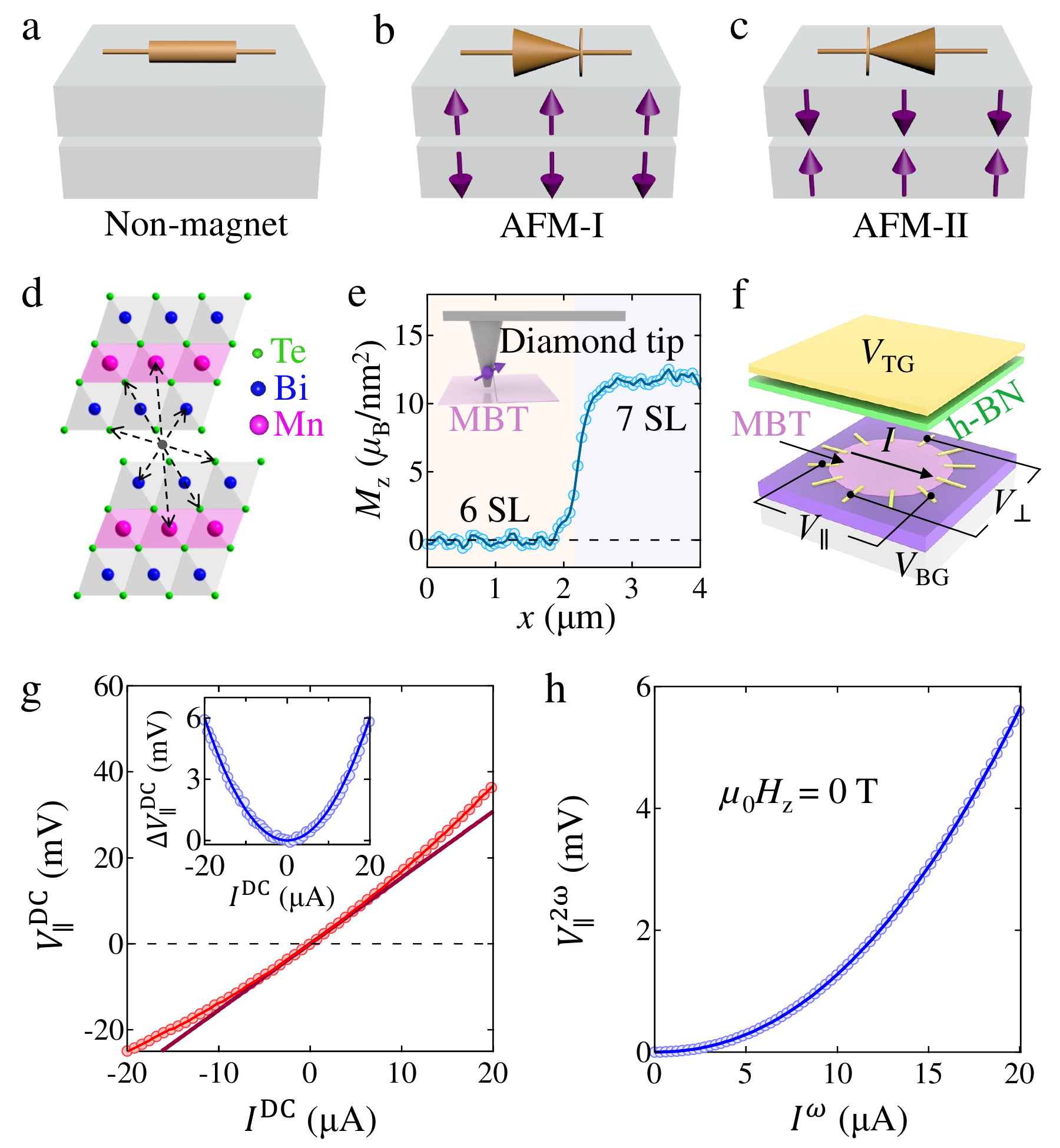}
\vspace{-0cm}
\caption{{\bf Introduction to the $\mathcal{PT}$-symmetric antiferromagnetism in even-layered MnBi$_2$Te$_4$.} \textbf{a-c,} Schematic illustration showing the AFM diode effect. The lattice is centrosymmetric. The AFM state breaks inversion symmetry, leading to the AFM diode effect. The two opposite AFM spin states (AFM-I and II) have opposite diode signal. \textbf{d,} Lattice of an even-layered MnBi$_2$Te$_4$ (bilayer used as an example). The lattice is centrosymmetric with the inversion center in-between two layers. \textbf{e,} NV measurement in a MnBi$_2$Te$_4$ flake. The magnetization in 6SL MnBi$_2$Te$_4$ is negligible. \textbf{f,} Schematic illustration of an even-layered MnBi$_2$Te$_4$ device with dual gating. \textbf{g,} DC $I-V$ characteristic of our 6SL MnBi$_2$Te$_4$ device. We subtract the measured voltage $V$ by a linear background, the resulting $\Delta V$ (shown by the inset) shows a quadratic behavior. \textbf{h,} AC $I-V$ characteristic of the 6SL MnBi$_2$Te$_4$ device. We inject a $\omega$ current $I^{\omega}$ and record the 2$\omega$ voltage $V^{2\omega}$.  }
\label{Fig1}
\end{figure*}

In this paper, we propose the concept of the antiferromagnetic diode effect and present its experimental observation, i.e., even in a centrosymmetric crystal without directional charge separation, the spins of an (AFM) can generate a directional current upon AC excitation (Fig.~\ref{Fig1}\textbf{b,c}). We also demonstrate important prototypical devices including the microwave harvester and the in-plane FET. We further demonstrate the electrical sum-frequency generation (SFG). The electrical SFG, which is analogous to the optical SFG previously discovered in wide-gap insulators, represents a powerful, novel tool to detect the nonlinear responses in quantum materials.

A key character is that the diode direction should be opposite for the two opposite AFM states (AFM-I and AFM-II shown in Fig.~\ref{Fig1}\textbf{b,c}). However, due to the absence of net magnetization (Fig.~\ref{Fig1}e and Extended Data Fig.~\ref{Compensated_AFM}), it is difficult to detect and control the two opposite AFM states, especially in a device setting. Moreover, many AFMs have multiple domains \cite{cheong2020seeing}. If domains of AFM-I and AFM-II have the same population, the AFM diode effect will cancel. Therefore, to test the AFM diode effect, we need an AFM material that can be prepared into a single domain. We also need a detection method that can distinguish AFM-I and AFM-II in a device setting.

The 2D even-layered MnBi$_2$Te$_4$, a prototypical $\mathcal{PT}$-symmetric AFM, is an ideal platform \cite{Otrokov2019a, Otrokov2019unique,Li2019a,Zhang2019a,Hao2019gapless,Zhang2020, Liu2020, Deng2020, Liu2020a,Trang2020, deng2021high, yang2021odd, Ovchinnikov2020,gao2021layer,cai2021electric, li2021nonlocal,tai2021polarity,Zhang2022diodic}. Its lattice is centrosymmetric (Fig.~\ref{Fig1}\textbf{d}) but its AFM spins break inversion symmetry (Fig.~\ref{Fig1}\textbf{b,c}). Moreover, our recent experiments \cite{Qiu2023axion} demonstrated a novel type of optical circular dichroism (the AFM CD), which can distinguish AFM-I and AFM-II even though both have zero net magnetization. Using this AFM CD, we found that the even-layered MnBi$_2$Te$_4$ indeed can be prepared in a single domain state (shown below).

\vspace{0.5cm}
\textbf{Observation of the antiferromagnetic diode effect: simultaneous optical and electrical measurements}

Overall, we have fabricated 13 devices with either the Hall bar electrodes or the radially distributed electrodes. They all give consistent results as a function of AFM order, spatial direction, doping, vertical electric field, etc. Here we focus on Device A, which is a 6SL MnBi$_2$Te$_4$ sample with radially distributed electrodes (Fig.~\ref{Fig1}\textbf{f}) and dual gating. The dual gating allows us to simultaneously control the carrier density $n_e$ and the vertical electric field $E_z$. All measurements were performed without a magnetic field at $T=1.7$ K unless otherwise noted. We start from the DC   longitudinal   $I-V$ characteristics. We observe a clear diode behavior as shown in Fig.~\ref{Fig1}\textbf{g}: a larger resistance in one direction but a smaller resistance in the opposite direction. We subtract our $I-V$ data with the linear background. The difference (inset of Fig.~\ref{Fig1}\textbf{g}) shows clear quadratic behavior $\Delta V=V-\alpha I=\beta I^2$. 
%Moreover, one can check that, in $V=\alpha I+\beta I^2$,  if we reverse the sign of $\beta$, the polarity of the diode behavior will flip. 
Therefore, a diode effect is characterized by the second-order nonlinear transport. Such nonlinear transport can be measured by the lock-in technique (Fig.~\ref{Fig1}\textbf{h}), where we inject an AC current $I^{\omega}$ and probe the second-harmonic voltage $V^{2\omega}$.

\begin{figure*}[t]
\includegraphics[width=11cm]{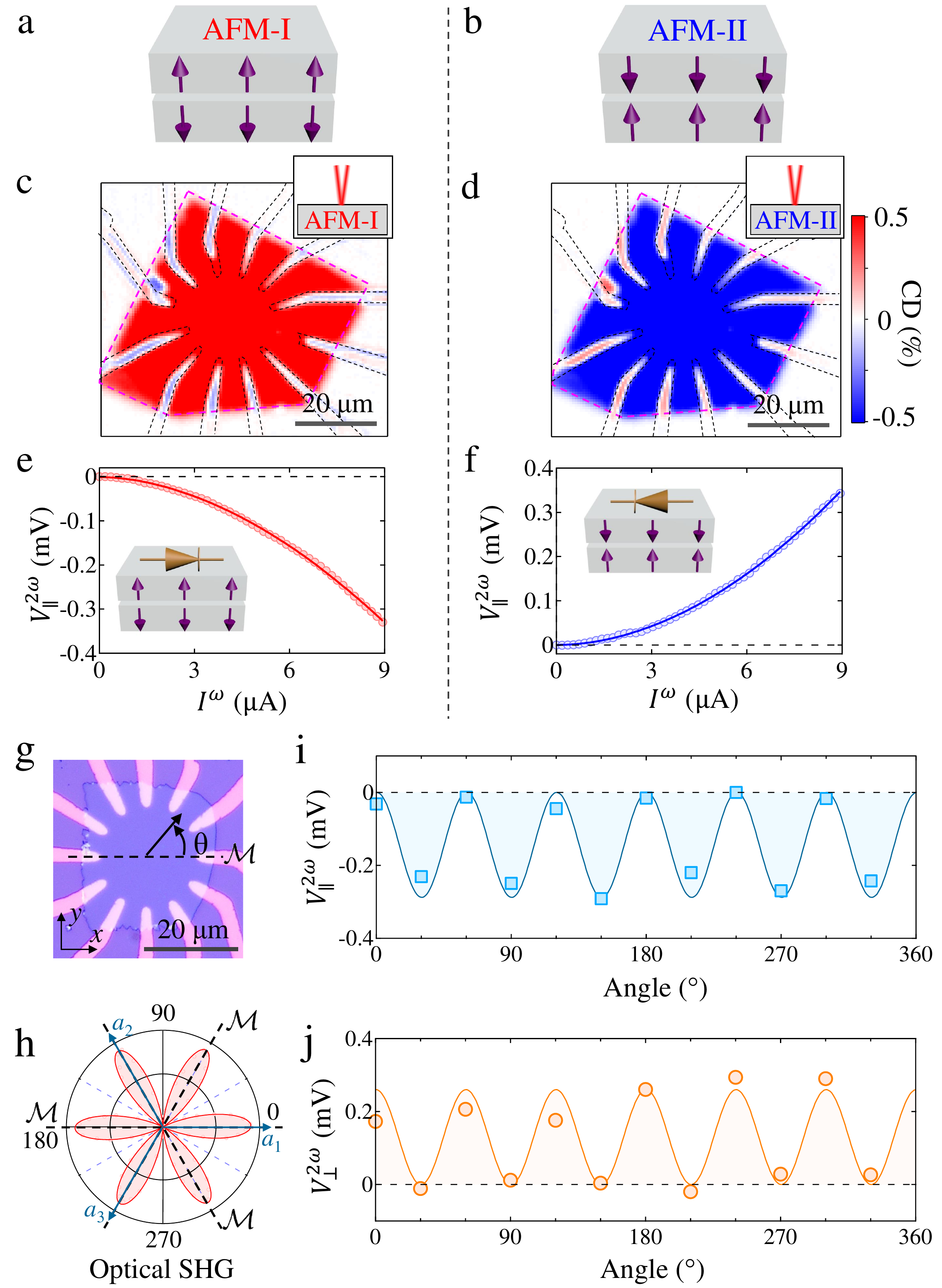}
\vspace{-0.5cm}
\caption{{\bf Observation of antiferromagnetic diode effect.} \textbf{a,} A schematic of particular AFM state (AFM-I) which can be prepared by sweeping the magnetic field under a fixed electric field following the procedures established by previous works \cite{gao2021layer} (see Extended Data Fig.~\ref{Linear_AFM}). \textbf{c,d,} As shown by our recent work \cite{Qiu2023axion}, a $\mathcal{PT}$-symmetric with zero net magnetization can support nonzero CD in the reflection channel (the AFM CD), whose sign is opposite for AFM-I and II. Therefore, our spatial mapping of the reflection CD here show that our 6SL-MnBi$_2$Te$_4$ device is prepared into a single-domain AFM state depending on the preparation procedure explained in Extended Data Fig.~\ref{Linear_AFM}. \textbf{e,f,} Nonlinear voltage as a function of incident current for AFM-I and II. \textbf{g,h,} In Device A, we determined the crystalline axes by the optical second-harmonic generation (panel h), and we deliberately align the electrodes with the crystalline axes (panel g). \textbf{i,j,} Angular dependence of longitudinal (panel i) and transverse (panel j) nonlinear voltage when the direction of the current is varied.  These data was taken at $I^{\omega}$ = 8 $\mu$A. }
\label{Fig2}
\end{figure*}

We now perform simultaneous optical and electrical measurements. Optically, we use the circular dichroism (CD) to confirm that our 6SL MnBi$_2$Te$_4$ is indeed in a single-domain state \cite{Qiu2023axion}. Then electrically, we detect the nonlinear transport signal to search for the AFM diode effect.  As demonstrated by previous experiments, the $\mathcal{PT}$-symmetric AFM order can be controlled by $\mathbf{E}\cdot\mathbf{B}$ field \cite{iyama2013magnetoelectric,jiang2018electric,gao2021layer}. We follow the established procedures \cite{gao2021layer}: under a fixed vertical electric field ($E_z=-0.2$ V/nm), we sweep $B_z$ from $-8$ T to 0 T to prepare AFM-I, and from $+8$ T to 0 T to prepare AFM-II (Extended Data Fig.~\ref{Linear_AFM}). To prove that we have indeed prepared a single-domain AFM state, we performed spatial-resolved CD measurements. As mentioned above, a $\mathcal{PT}$-symmetric AFM with zero net magnetization can support the AFM CD in the reflection channel \cite{Krichevtsov1993spontaneous, ahn2022theory, Qiu2023axion}, whose signal is opposite between AFM-I and AFM-II. As shown in Fig.~\ref{Fig2}\textbf{c,d}, the entire sample has the same CD signal. In such single-domain AFM states, we now measure their longitudinal nonlinear transport signals. Our data (Fig.~\ref{Fig2}\textbf{e,f} and Extended Data Fig.~\ref{hysB_Nonlinear}) show that the two AFM states feature opposite longitudinal nonlinear voltage $V^{2\omega}_{\|}$, hence the opposite diode effect. In contrast, the longitudinal linear voltage (i.e., the regular resistance) shown in Extended Data Fig.~\ref{Linear_AFM} is the same for the two AFM states. We further measure $\sigma^{2\omega}$ while increasing the temperature. Shown in Extended Data Fig.~\ref{Scaling}, the nonlinear signal disappears above the N\'eel temperature $T_{\textrm{N}}\simeq21$ K. 

Therefore, our data show that the single-domain AFM state in 6SL MnBi$_2$Te$_4$ exhibits a clear nonlinear transport signal. Its sign depends on the AFM state. Hence, the observed nonlinear effect directly arises from the inversion breaking of the AFM spin structure, which is opposite for AFM-I and AFM-II. This finding is equivalent to that the nonlinear conductivity $\sigma^{\textrm{NL}}_{ijk}$ is odd under time-reversal operation, a hallmark of the AFM diode effect ($\sigma^{\textrm{NL}}_{ijk}=\frac{J_i^{2\omega}}{E_j^{\omega}E_k^{\omega}}=\frac{V^{2\omega}_i}{I_j^{\omega}I_k^{\omega}R^3}\frac{l^2}{wd}$ ($R$ is sample resistance and $l,w,d$ are sample length, width, and thickness)).
 
\vspace{0.5cm}
\textbf{Systematic studies of the intrinsic AFM diode effect}

To further confirm that the nonlinear signal is an intrinsic effect of the AFM, we study how the nonlinear signal depends on the crystallographic directions. In Device A, we deliberately align the electrodes with the crystalline axes (Fig.~\ref{Fig2}\textbf{g}) which were determined by the optical second-harmonic generation (Fig.~\ref{Fig2}\textbf{h}). These preparations allow us to test the crystalline direction dependence of the nonlinear signal. The longitudinal nonlinear signal oscillates between zero and nonzero, respecting the three-fold rotational symmetry of the MnBi$_2$Te$_4$. In addition to the longitudinal nonlinear signal, we actually also detect a transverse nonlinear signal, as shown in Fig.~\ref{Fig2}\textbf{j}. In the latter part of the paper (Fig.~\ref{Fig4}), we will study the exchange property to differentiate this transverse diode signal from a nonlinear Hall effect. Here, we focus on the crystallographic directional dependence. By comparing Figs.~\ref{Fig2}\textbf{i,j}, we can observe the following:  the longitudinal $V^{2\omega}_{\parallel}$ is zero when the current is parallel to a crystalline axis (dotted lines in Fig.~\ref{Fig2}\textbf{h}) but maximum when the current is perpendicular. By contrast, the transverse $V^{2\omega}_{\perp}$ has the opposite directional dependence: $V^{2\omega}_{\perp}$ is maximum when the current is parallel to a crystalline axis but zero when the current is perpendicular (see also Extended Data Fig.~\ref{NLV_Direction}).

Such directional dependence can help us understand the nature of the observed nonlinear signals (intrinsic obeying the symmetry of the AFM order or artifact due to device imperfection/asymmetry). If it is an intrinsic response of the AFM state of even-layered MnBi$_2$Te$_4$, the nonlinear signals and their directional dependence should be consistent with the symmetry of the AFM state, which is $-3'm'$ for even-layered MnBi$_2$Te$_4$. Through symmetry analysis, one can show that the group $-3'm'$ dictates that $\sigma^{\textrm{NL}}_{yxx}=-\sigma^{\textrm{NL}}_{yyy}$ are nonzero but $\sigma^{\textrm{NL}}_{xxx}=\sigma^{\textrm{NL}}_{xyy}=0$, consistent with our data, which therefore strongly suggests that our observed nonlinear signals is intrinsic.

\begin{figure*}[t]
\includegraphics[width=12cm]{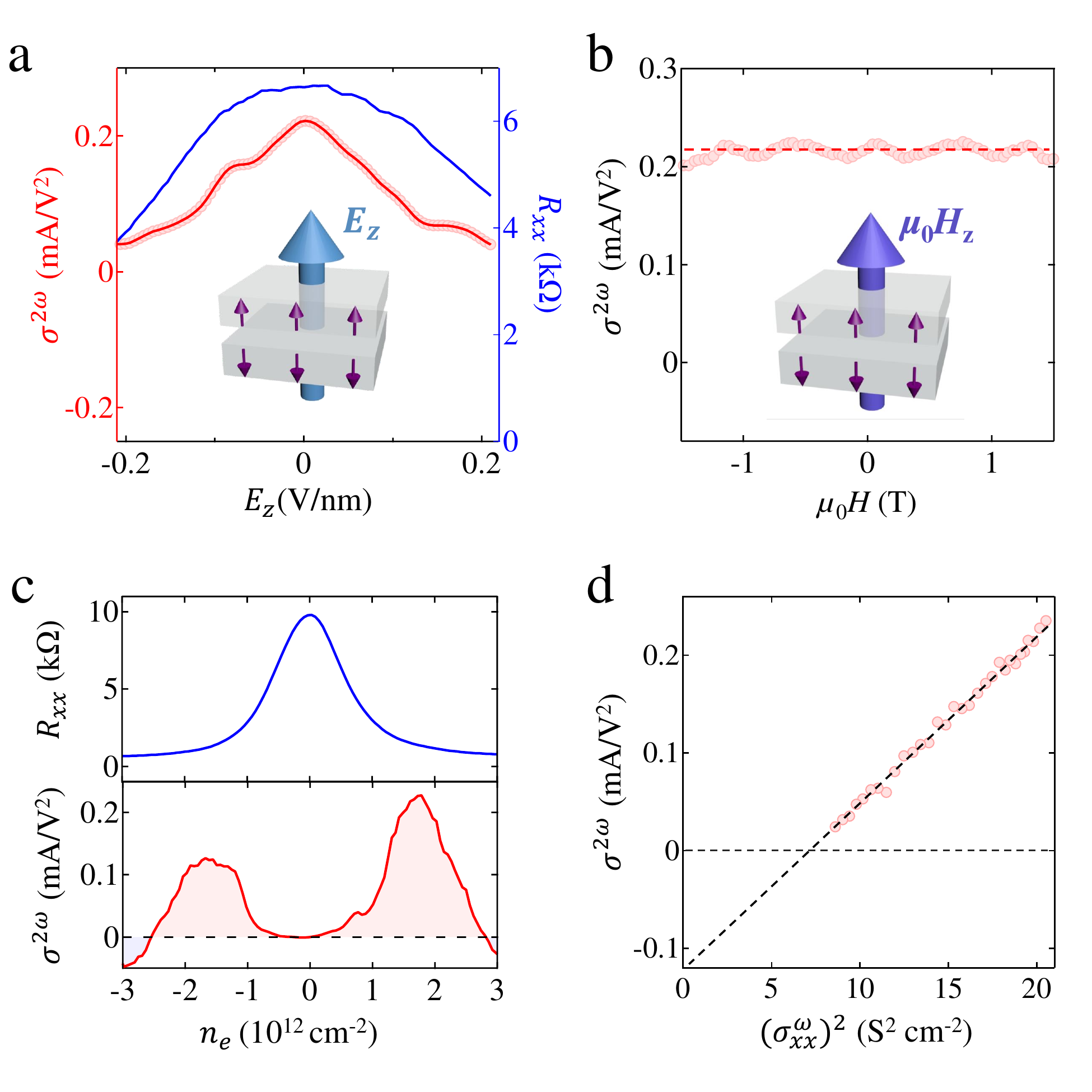}
\vspace{-0.5cm}
\caption{{\bf Systematic investigations of the intrinsic antiferromagnetic diode effect.}  \textbf{a,} Nonlinear conductivity and longitudinal resistance as a function of vertical electric field $E_z$. \textbf{b,} Nonlinear conductivity as a function of vertical magnetic field $B_z$. \textbf{c,} Experimentally measured longitudinal resistance (upper panel) and nonlinear conductivity (lower panel) as a function of charge density $n_e$. \textbf{d,} The scaling between the nonlinear conductivity and the square of regular linear conductivity. The dark dashed line is the linear fit to the scaling curve. }
\label{Fig3}
\end{figure*}

\begin{figure*}[t]
\includegraphics[width=14cm]{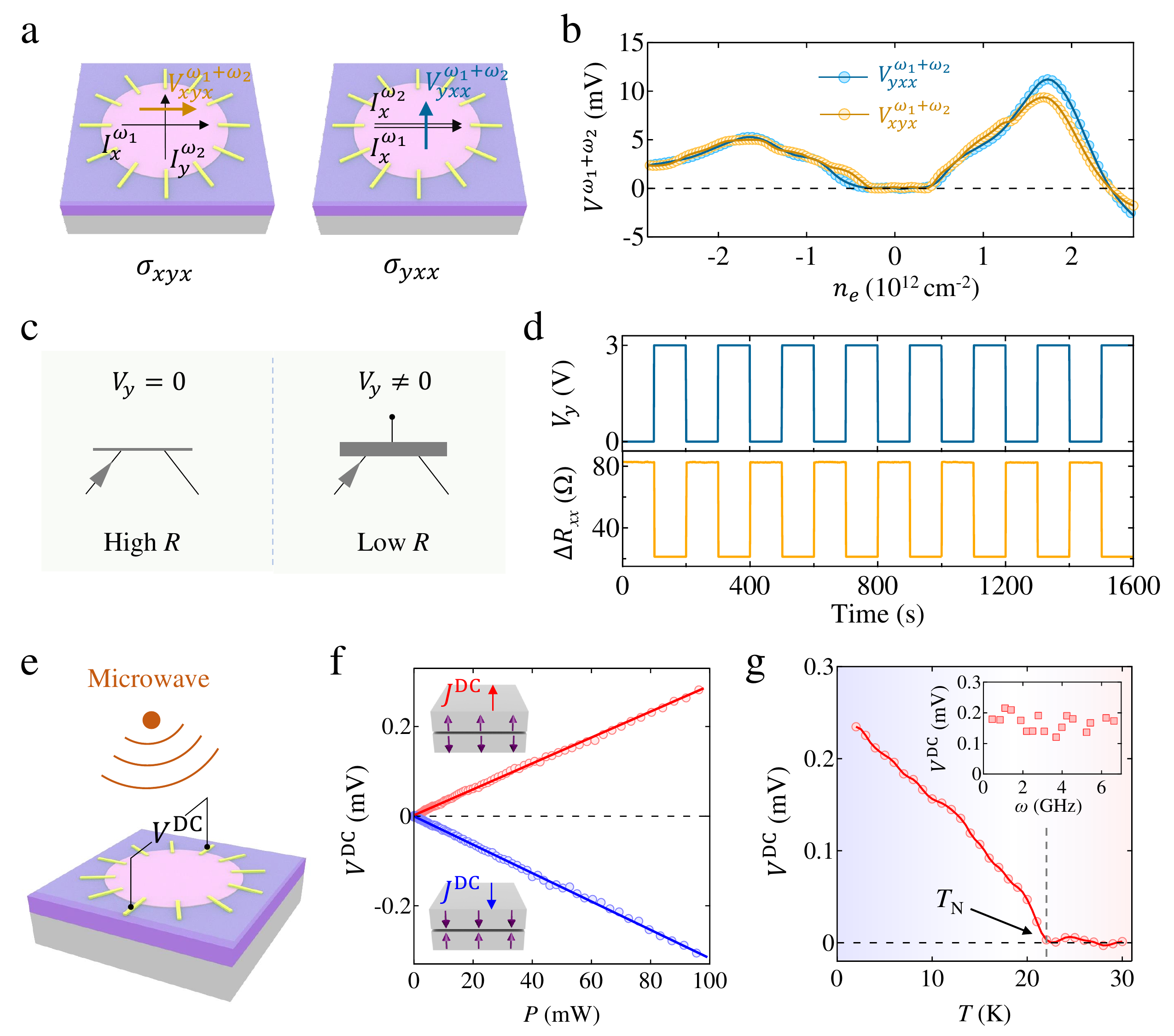}
\vspace{-0.2cm}
\caption{{\bf Sum-frequency generation (SFG) measurements and applications of the AFM diode effect.} \textbf{a,} The schematic of SFG measurements. We inject two currents of different frequencies ($I^{\omega_1}$ and $I^{\omega_2}$) and probe the SFG voltage $V^{\omega_1+\omega_2}$ (see Methods and Extended Data Fig.~\ref{SFG} for technical details). This electrical SFG is particularly advantageous to study $\sigma^{\textrm{NL}}_{xyx}$: The fact that the two currents have different frequencies makes it feasible to inject them along different directions. \textbf{b,}  $V^{\textrm{NL}}_{xyx}$ and $V^{\textrm{NL}}_{yxx}$ measured using SFG measurement method. The SFG signal is symmetric upon exchanging the first two indices of $V^{\textrm{ND}}_{ijk}$, which is consistent with the AFM group $-3'm'$ of MnBi$_2$Te$_4$. \textbf{c,} The SFG realizes an intrinsic in-plane field effect. When we take $\omega_2\rightarrow 0$, the SFG experiment becomes $J_x^{\omega_1}=\sigma^{\textrm{NL}}_{xyx}E^{\textrm{DC}}_yE^{\omega_1}_x $. This equation can be further rewritten as ${\Delta \sigma_{xx}}\propto\sigma^{\textrm{NL}}_{xyx}E^{\textrm{DC}}_y$. In other words, the conductance (resistance) along $x$ can be directly modulated by an electric field bias along $y$. \textbf{d,} Measured change of resistance $R_{xx}$ ($\Delta R_{xx}=V^{\omega_1}_x/I^{\omega_1}_x$) as a function of an electric field bias along $V_y$. \textbf{e,} Schematic illustration of the wireless microwave radiation harvesting. \textbf{f,} Measured DC rectification voltage as a function of the wireless electromagnetic power for both AFM states. \textbf{g,} Measured DC rectification voltage as a function of temperature. The inset shows the DC rectification voltage as a function of frequency $\omega$.}
\label{Fig4}
\end{figure*}

We further perform systematic studies. Figure~\ref{Fig3}\textbf{a} shows the $E_z$ dependence of the nonlinear conductivity $\sigma^{2\omega}$, from which we see that $V^{2\omega}$ in our MnBi$_2$Te$_4$ system does not require a finite electric $E_z$ field. Because $E_z$ leads to a finite electric polarization $P_z=\chi_e E_z$, equivalently, our system supports nonlinear transport even though it has zero electric polarization ($P_z=0$). Figure~\ref{Fig3}\textbf{b} shows the $B_z$ dependence of the $V^{2\omega}$, from which we see that $V^{2\omega}$ in our system also does not require a finite magnetic $B_z$ field. Therefore, the observed nonlinear signal is not induced by external $E$ or $B$ field. Figure~\ref{Fig3}\textbf{c} shows the charge density $n_e$ dependence of $\sigma^{2\omega}$. In order to uncover further information about the microscopic mechanisms, we now studied the relaxation time $\tau$ dependence of $\sigma^{2\omega}$ by measuring the scaling between the $\sigma^{2\omega}$ and the linear Drude conductivity $\sigma_{xx}^\omega$ (through temperature dependence). Our data (Fig.~\ref{Fig3}\textbf{d} and Extended Data Fig.~\ref{Scaling}) show that the nonlinear conductivity $\sigma^{2\omega}$ has two components, one proportional to $\tau^2$ and the other proportional to $\tau^0$ ($\sigma^{2\omega} = a _0\tau^0+a_2\tau^2$). The existence of two components suggests multiple, coexisting microscopic mechanisms. Based on theoretical studies of nonlinear transport in AFMs \cite{gao2014field, Wang2021Intrinsic,liu2021intrinsic, holder2021mixed, lahiri2022intrinsic, Smith2022momentum, ma2022anomalous, hayami2022nonlinear, kaplan2022unification,huang2023nonlinear, huang2023Scaling, atencia2023disorder}, we propose that the $\tau^2$ component can arise from the nonlinear Drude effect and disorder scattering \cite{ma2022anomalous, huang2023Scaling, atencia2023disorder}, whereas the $\tau^0$ component may arise from the quantum metric and disorder scattering \cite{huang2023Scaling, atencia2023disorder}. We have directly computed the nonlinear Drude conductivity based on the intrinsic band structure of 6SL MnBi$_2$Te$_4$ (Extended Data Fig.~\ref{Band_AFM}\textbf{c}). Their difference suggests that beyond quantum metric, quantum metric and disorder scattering can also have important contribution to our data. We note that, in contrast to the well-known nonlinear Drude effect, the quantum metric and disorder scattering are still under active theoretical investigation \cite{gao2014field,Wang2021Intrinsic,liu2021intrinsic, holder2021mixed, lahiri2022intrinsic, Smith2022momentum, ma2022anomalous, hayami2022nonlinear, kaplan2022unification,huang2023nonlinear, huang2023Scaling}. Our current data is insufficient to isolate the quantum metric contribution from disorder scattering induced contributions. In the future, it would be interesting to measure the nonlinear conductivity of pristine MnBi$_2$Te$_4$ at terahertz frequencies ($\omega\tau\gg1$), where the disorder scattering may be suppressed \cite{kovalev2010anomalous, fujimoto2023observation}. 

\vspace{0.5cm}
\textbf{Electrical sum-frequency generation by AFM spins}

 We now report a novel electrical sum frequency generation (SFG) in our AFM MnBi2Te4 devices. While the optical SFG has been well-known, electrical SFG has been rarely studied. We show that the electrical SHG can study the exchange property to differentiate our transverse diode signal from a nonlinear Hall effect. To motivate the importance of the exchange property, we first give an example of the linear conductivity, which is a simple $2\times2$ tensor  $\left(\begin{smallmatrix}
        \sigma_{xx} & \sigma_{xy} \\
        \sigma_{yx} & \sigma_{yy} \\
    \end{smallmatrix}\right)$ in 2D. There are two distinctly different physical mechanisms for a nonzero transverse linear conductance: (1) the Hall effect and (2) the crystalline anisotropy \cite{walmsley2017determination}. These two distinctly different kinds of transverse linear conductance are differentiated by their exchange properties. The Hall effect is anti-symmetric, i.e., $\sigma_{xy}=-\sigma_{yx}$. By contrast, the anisotropy induced transverse conductance is symmetric $\sigma_{xy}=\sigma_{yx}$ (Extended Data Figs.~\ref{Anisotropy_Hall}\textbf{a,b}). %Therefore, the exchange property of the conductivity tensor is crucial in differentiating its origin.

The same logic can be applied to the nonlinear conductivity. Our observation of a transverse nonlinear voltage means that $\sigma^{\textrm{NL}}_{yxx}$ is nonzero. However, to understand its origin, we should study the exchange property (Extended Data Figs.~\ref{Anisotropy_Hall}\textbf{c,d}), i.e., to compare $\sigma^{\textrm{NL}}_{yxx}$ and $\sigma^{\textrm{NL}}_{xyx}$. If $\sigma^{\textrm{NL}}_{yxx}=-\sigma^{\textrm{NL}}_{xyx}$ (anti-symmetric), then the transverse nonlinear signal is a nonlinear Hall effect \cite{tsirkin2022separation}. By contrast, if $\sigma^{\textrm{NL}}_{yxx}=\sigma^{\textrm{NL}}_{xyx}$ (symmetric), then it is not a Hall effect. Rather, it is an AFM diode effect just like the longitudinal component.

Measuring $\sigma^{\textrm{NL}}_{xyx}$ in an electrical second-harmonic generation experiment, however, is difficult, because it requires passing one current $I^{\omega}$ along $x$ and another $I^{\omega}$ along $y$ without undesired interference. As such, in order to isolate and study $\sigma^{\textrm{NL}}_{xyx}$, we present a novel measurement that was unachieved previously, the electrical sum-frequency generation (SFG). We inject two currents of different frequencies ($I^{\omega_1}$ and $I^{\omega_2}$) and probe the SFG voltage $V^{\omega_1+\omega_2}$. (We need to set $\omega_1 \gg \omega_2$ to achieve the electrical SFG measurements. See Methods for technical details). This new electrical SFG is particularly advantageous to study $\sigma^{\textrm{NL}}_{xyx}$: The fact that the two currents have different frequencies makes it feasible to clearly separate their trajectories along orthogonal directions (see Extended Data Fig.~\ref{SFG} and the Methods). As shown in Fig.~\ref{Fig4}\textbf{b}, we show that our sample indeed supports strong SFG $V^{\omega_1+\omega_2}_{xyx}$ by injecting $I^{\omega_1}//x$ and $I^{\omega_2}//y$. Moreover, we use this SFG method to measure both $\sigma^{\textrm{NL}}_{xyx}$ and $\sigma^{\textrm{NL}}_{yxx}$. Our data shows that $\sigma^{\textrm{NL}}_{xyx}=\sigma^{\textrm{NL}}_{yxx}$,  which shows that our transverse nonlinear signal is not a Hall effect. Rather, it is an AFM diode effect just like the longitudinal component. More broadly, our SFG measurement establishes a new, universal method to study the nonlinear electrical transport processes of quantum materials, which can be applied to many other systems.

\vspace{0.5cm}
\textbf{In-plane field effect transistor and AFM harvesting of wireless radiation}

Interestingly, the $\sigma^{\textrm{NL}}_{xyx}$ realizes a novel, intrinsic in-plane field effect transistor (Fig.~\ref{Fig4}\textbf{c}  and Extended Data Fig.~\ref{Inplane_Field_Temp}) and potentially an electro-optical modulator. To see this, let us take $\omega_2\rightarrow 0$. Then the SFG experiment evolves into $\Delta J_x^{\omega_1}=\sigma^{\textrm{NL}}_{xyx}E^{\textrm{DC}}_y E^{\omega_1}_x={\Delta \sigma_{xx}}E^{\omega_1}_x$. Hence we have ${\Delta \sigma_{xx}}=\sigma^{\textrm{NL}}_{xyx}E^{\textrm{DC}}_y$. In other words, the conductance (resistance) along $x$ can be directly modulated by an electric field bias along $y$. In Fig.~\ref{Fig4}\textbf{c,d}, we apply an AC bias along $x$ and a DC bias along $y$. By turning on and off the DC bias along $y$, we indeed observe a clear change of resistance along $x$, therefore demonstrating the in-plane field effect. In traditional field effect transistors, the electric field is supplied by the gate along the out-of-plane direction. In many cases, the capacitance of gate dielectrics is the bottleneck limiting its on/off speed. By contrast, in our case, gating is not needed, the electric field is supplied by a bias across two contacts along the in-plane direction. This in-plane field effect can potentially enable ultrafast transistors. 
Further increasing $\omega_1$ to terahertz can potentially realize an electro-optical modulator, where the optical conductivity (the dielectric constant) of the terahertz light can be modulated by the DC bias $E^{\textrm{DC}}_y$. Whereas known electro-optic effect occurs in insulators, the effect envisioned here arises from a Fermi surface contribution.

In addition to the second-harmonic generation and the sum-frequency generation, the second-order transport also enables the rectification. We demonstrate the rectification of wireless microwave radiation into DC electricity based on the AFM diode effect. As shown in Fig.~\ref{Fig4}\textbf{e}, wireless microwaves were irradiated onto the device from free space (see Methods). A DC electric voltage linear to the microwave power (hence quadratic to the microwave electric field) is observed (Fig.~\ref{Fig4}\textbf{f}). Moreover, the DC signal can be directly switched by reversing the AFM order (Fig.~\ref{Fig4}\textbf{f}). The rectification shows a broadband response (inset of Fig.~\ref{Fig4}\textbf{g}), including the WiFi frequencies ($2.4$ GHz and $5$ GHz). As such, our data demonstrates a wireless rectification device that can be controlled by the AFM spins.

In summary, our work demonstrates the intrinsic spin-induced nonlinearity, which represents a novel, universal character for all $\mathcal{PT}$-symmetric AFM conductors. This is achieved by (1) Spatially-resolved optical method (the AFM circular dichroism) shows that we indeed achieve single AFM domain; (2) Electrical SFG measurements demonstrated that the nonlinear conductivity we studied in pristine is  MnBi$_2$Te$_4$ indeed symmetric. (3) Systematic dependences as a function of temperature, spatial direction, AFM domain, $E$-field and $B$-field are all consistent with the intrinsic magnetic group of the AFM order of MnBi$_2$Te$_4$ ($-3'm'$). Moreover, we developed and demonstrated the electrical sum-frequency generation measurements, which is analogous to the optical SFG previously discovered in wide-gap insulators. This electrical SFG represents a powerful, conceptually novel tool to detect the nonlinear responses in quantum materials. For strongly correlated materials, our AFM diode effect provides a sensitive probe of the exotic $\mathcal{PT}$-symmetric order parameter of the Fermi surface electrons such as the loop current (pseudo-gap) states \cite{orenstein2011optical, varma2014gyrotropic,Zhang2022diodic}. Beyond fundamental materials research, our work here implies new device concepts, such as the AFM logic circuits, AFM-based microwave harvester, and self-powered AFM spintronic devices. In the future, the identification of room-temperature AFM materials with large nonlinear responses can lead to possibilities that combine AFM spintronics, nonlinear electronics and topological materials.

\vspace{0.5cm}

\bibliographystyle{naturemag}
\bibliography{Diode_03122024}

\begin{thebibliography}{10}
\expandafter\ifx\csname url\endcsname\relax
  \def\url#1{\texttt{#1}}\fi
\expandafter\ifx\csname urlprefix\endcsname\relax\def\urlprefix{URL }\fi
\providecommand{\bibinfo}[2]{#2}
\providecommand{\eprint}[2][]{\url{#2}}

\bibitem{lee2006doping}
\bibinfo{author}{Lee, P.~A.}, \bibinfo{author}{Nagaosa, N.} \& \bibinfo{author}{Wen, X.-G.}
\newblock \bibinfo{title}{Doping a {M}ott insulator: Physics of high-temperature superconductivity}.
\newblock \emph{\bibinfo{journal}{Rev. Mod. Phys.}} \textbf{\bibinfo{volume}{78}}, \bibinfo{pages}{17--84} (\bibinfo{year}{2006}).

\bibitem{mong2010antiferromagnetic}
\bibinfo{author}{Mong, R.~S.}, \bibinfo{author}{Essin, A.~M.} \& \bibinfo{author}{Moore, J.~E.}
\newblock \bibinfo{title}{Antiferromagnetic topological insulators}.
\newblock \emph{\bibinfo{journal}{Phys. Rev. B}} \textbf{\bibinfo{volume}{81}}, \bibinfo{pages}{245209} (\bibinfo{year}{2010}).

\bibitem{tang2016dirac}
\bibinfo{author}{Tang, P.}, \bibinfo{author}{Zhou, Q.}, \bibinfo{author}{Xu, G.} \& \bibinfo{author}{Zhang, S.-C.}
\newblock \bibinfo{title}{Dirac fermions in an antiferromagnetic semimetal}.
\newblock \emph{\bibinfo{journal}{Nat. Phys.}} \textbf{\bibinfo{volume}{12}}, \bibinfo{pages}{1100--1104} (\bibinfo{year}{2016}).

\bibitem{qi2008topological}
\bibinfo{author}{Qi, X.-L.}, \bibinfo{author}{Hughes, T.~L.} \& \bibinfo{author}{Zhang, S.-C.}
\newblock \bibinfo{title}{Topological field theory of time-reversal invariant insulators}.
\newblock \emph{\bibinfo{journal}{Phys. Rev. B}} \textbf{\bibinfo{volume}{78}}, \bibinfo{pages}{195424} (\bibinfo{year}{2008}).

\bibitem{essin2009magnetoelectric}
\bibinfo{author}{Essin, A.~M.}, \bibinfo{author}{Moore, J.~E.} \& \bibinfo{author}{Vanderbilt, D.}
\newblock \bibinfo{title}{Magnetoelectric polarizability and axion electrodynamics in crystalline insulators}.
\newblock \emph{\bibinfo{journal}{Phys. Rev. Lett.}} \textbf{\bibinfo{volume}{102}}, \bibinfo{pages}{146805} (\bibinfo{year}{2009}).

\bibitem{jungwirth2016antiferromagnetic}
\bibinfo{author}{Jungwirth, T.}, \bibinfo{author}{Marti, X.}, \bibinfo{author}{Wadley, P.} \& \bibinfo{author}{Wunderlich, J.}
\newblock \bibinfo{title}{Antiferromagnetic spintronics}.
\newblock \emph{\bibinfo{journal}{Nat. Nanotechnol.}} \textbf{\bibinfo{volume}{11}}, \bibinfo{pages}{231--241} (\bibinfo{year}{2016}).

\bibitem{rikken2001electrical}
\bibinfo{author}{Rikken, G.}, \bibinfo{author}{F{\"o}lling, J.} \& \bibinfo{author}{Wyder, P.}
\newblock \bibinfo{title}{Electrical magnetochiral anisotropy}.
\newblock \emph{\bibinfo{journal}{Phys. Rev. Lett.}} \textbf{\bibinfo{volume}{87}}, \bibinfo{pages}{236602} (\bibinfo{year}{2001}).

\bibitem{rikken2005magnetoelectric}
\bibinfo{author}{Rikken, G.} \& \bibinfo{author}{Wyder, P.}
\newblock \bibinfo{title}{Magnetoelectric anisotropy in diffusive transport}.
\newblock \emph{\bibinfo{journal}{Phys. Rev. Lett.}} \textbf{\bibinfo{volume}{94}}, \bibinfo{pages}{016601} (\bibinfo{year}{2005}).

\bibitem{tokura2018nonreciprocal}
\bibinfo{author}{Tokura, Y.} \& \bibinfo{author}{Nagaosa, N.}
\newblock \bibinfo{title}{Nonreciprocal responses from non-centrosymmetric quantum materials}.
\newblock \emph{\bibinfo{journal}{Nat. Commun.}} \textbf{\bibinfo{volume}{9}}, \bibinfo{pages}{3740} (\bibinfo{year}{2018}).

\bibitem{ma2021topology}
\bibinfo{author}{Ma, Q.}, \bibinfo{author}{Grushin, A.~G.} \& \bibinfo{author}{Burch, K.~S.}
\newblock \bibinfo{title}{Topology and geometry under the nonlinear electromagnetic spotlight}.
\newblock \emph{\bibinfo{journal}{Nat. Mater.}} \textbf{\bibinfo{volume}{20}}, \bibinfo{pages}{1601--1614} (\bibinfo{year}{2021}).

\bibitem{orenstein2021topology}
\bibinfo{author}{Orenstein, J.} \emph{et~al.}
\newblock \bibinfo{title}{Topology and symmetry of quantum materials via nonlinear optical responses}.
\newblock \emph{\bibinfo{journal}{Annu. Rev. Condens. Matter Phys.}} \textbf{\bibinfo{volume}{12}}, \bibinfo{pages}{247--272} (\bibinfo{year}{2021}).

\bibitem{moore2010confinement}
\bibinfo{author}{Moore, J.~E.} \& \bibinfo{author}{Orenstein, J.}
\newblock \bibinfo{title}{Confinement-induced {B}erry phase and helicity-dependent photocurrents}.
\newblock \emph{\bibinfo{journal}{Phys. Rev. Lett.}} \textbf{\bibinfo{volume}{105}}, \bibinfo{pages}{026805} (\bibinfo{year}{2010}).

\bibitem{gao2014field}
\bibinfo{author}{Gao, Y.}, \bibinfo{author}{Yang, S.~A.} \& \bibinfo{author}{Niu, Q.}
\newblock \bibinfo{title}{Field induced positional shift of {B}loch electrons and its dynamical implications}.
\newblock \emph{\bibinfo{journal}{Phys. Rev. Lett.}} \textbf{\bibinfo{volume}{112}}, \bibinfo{pages}{166601} (\bibinfo{year}{2014}).

\bibitem{sodemann2015quantum}
\bibinfo{author}{Sodemann, I.} \& \bibinfo{author}{Fu, L.}
\newblock \bibinfo{title}{Quantum nonlinear {H}all effect induced by {B}erry curvature dipole in time-reversal invariant materials}.
\newblock \emph{\bibinfo{journal}{Phys. Rev. Lett.}} \textbf{\bibinfo{volume}{115}}, \bibinfo{pages}{216806} (\bibinfo{year}{2015}).

\bibitem{rectification}
\bibinfo{author}{Ideue, T.} \emph{et~al.}
\newblock \bibinfo{title}{Bulk rectification effect in a polar semiconductor}.
\newblock \emph{\bibinfo{journal}{Nat. Phys.}} \textbf{\bibinfo{volume}{13}}, \bibinfo{pages}{578--583} (\bibinfo{year}{2017}).

\bibitem{oh2016field}
\bibinfo{author}{Oh, Y.-W.} \emph{et~al.}
\newblock \bibinfo{title}{Field-free switching of perpendicular magnetization through spin--orbit torque in antiferromagnet/ferromagnet/oxide structures}.
\newblock \emph{\bibinfo{journal}{Nat. Nanotechnol.}} \textbf{\bibinfo{volume}{11}}, \bibinfo{pages}{878--884} (\bibinfo{year}{2016}).

\bibitem{yasuda2016large}
\bibinfo{author}{Yasuda, K.} \emph{et~al.}
\newblock \bibinfo{title}{Large unidirectional magnetoresistance in a magnetic topological insulator}.
\newblock \emph{\bibinfo{journal}{Phys. Rev. Lett.}} \textbf{\bibinfo{volume}{117}}, \bibinfo{pages}{127202} (\bibinfo{year}{2016}).

\bibitem{he2018bilinear}
\bibinfo{author}{He, P.} \emph{et~al.}
\newblock \bibinfo{title}{Bilinear magnetoelectric resistance as a probe of three-dimensional spin texture in topological surface states}.
\newblock \emph{\bibinfo{journal}{Nat. Phys.}} \textbf{\bibinfo{volume}{14}}, \bibinfo{pages}{495--499} (\bibinfo{year}{2018}).

\bibitem{godinho2018electrically}
\bibinfo{author}{Godinho, J.} \emph{et~al.}
\newblock \bibinfo{title}{Electrically induced and detected {N}\'eel vector reversal in a collinear antiferromagnet}.
\newblock \emph{\bibinfo{journal}{Nat. Commun.}} \textbf{\bibinfo{volume}{9}}, \bibinfo{pages}{4686} (\bibinfo{year}{2018}).

\bibitem{ma2019observation}
\bibinfo{author}{Ma, Q.} \emph{et~al.}
\newblock \bibinfo{title}{Observation of the nonlinear {H}all effect under time-reversal-symmetric conditions}.
\newblock \emph{\bibinfo{journal}{Nature}} \textbf{\bibinfo{volume}{565}}, \bibinfo{pages}{337--342} (\bibinfo{year}{2019}).

\bibitem{Kang2019nonlinear}
\bibinfo{author}{Kang, K.}, \bibinfo{author}{Li, T.}, \bibinfo{author}{Sohn, E.}, \bibinfo{author}{Shan, J.} \& \bibinfo{author}{Mak, K.~F.}
\newblock \bibinfo{title}{Nonlinear anomalous {H}all effect in few-layer {W}{T}e$_2$}.
\newblock \emph{\bibinfo{journal}{Nat. Mater.}} \textbf{\bibinfo{volume}{18}}, \bibinfo{pages}{324--328} (\bibinfo{year}{2019}).

\bibitem{kumar2021room}
\bibinfo{author}{Kumar, D.} \emph{et~al.}
\newblock \bibinfo{title}{Room-temperature nonlinear {H}all effect and wireless radiofrequency rectification in {W}eyl semimetal {T}a{I}r{T}e$_4$}.
\newblock \emph{\bibinfo{journal}{Nat. Nanotechnol.}} \textbf{\bibinfo{volume}{16}}, \bibinfo{pages}{421--425} (\bibinfo{year}{2021}).

\bibitem{sinha2022berry}
\bibinfo{author}{Sinha, S.} \emph{et~al.}
\newblock \bibinfo{title}{Berry curvature dipole senses topological transition in a moir\'e superlattice}.
\newblock \emph{\bibinfo{journal}{Nat. Phys.}} \textbf{\bibinfo{volume}{18}}, \bibinfo{pages}{765--770} (\bibinfo{year}{2022}).

\bibitem{isobe2020high}
\bibinfo{author}{Isobe, H.}, \bibinfo{author}{Xu, S.-Y.} \& \bibinfo{author}{Fu, L.}
\newblock \bibinfo{title}{High-frequency rectification via chiral {B}loch electrons}.
\newblock \emph{\bibinfo{journal}{Sci. Adv.}} \textbf{\bibinfo{volume}{6}}, \bibinfo{pages}{eaay2497} (\bibinfo{year}{2020}).

\bibitem{yasuda2020large}
\bibinfo{author}{Yasuda, K.} \emph{et~al.}
\newblock \bibinfo{title}{Large non-reciprocal charge transport mediated by quantum anomalous {H}all edge states}.
\newblock \emph{\bibinfo{journal}{Nat. Nanotechnol.}} \textbf{\bibinfo{volume}{15}}, \bibinfo{pages}{831--835} (\bibinfo{year}{2020}).

\bibitem{zhao2020magnetic}
\bibinfo{author}{Zhao, W.} \emph{et~al.}
\newblock \bibinfo{title}{Magnetic proximity and nonreciprocal current switching in a monolayer {WT}e$_2$ helical edge}.
\newblock \emph{\bibinfo{journal}{Nat. Mater.}} \textbf{\bibinfo{volume}{19}}, \bibinfo{pages}{503--507} (\bibinfo{year}{2020}).

\bibitem{tsirkin2022separation}
\bibinfo{author}{Tsirkin, S.} \& \bibinfo{author}{Souza, I.}
\newblock \bibinfo{title}{On the separation of hall and ohmic nonlinear responses}.
\newblock \emph{\bibinfo{journal}{SciPost Phys. Core}} \textbf{\bibinfo{volume}{5}}, \bibinfo{pages}{039} (\bibinfo{year}{2022}).

\bibitem{Lai2021third}
\bibinfo{author}{Lai, S.} \emph{et~al.}
\newblock \bibinfo{title}{Third-order nonlinear {H}all effect induced by the {B}erry-connection polarizability tensor}.
\newblock \emph{\bibinfo{journal}{Nat. Nanotechnol.}} \textbf{\bibinfo{volume}{16}}, \bibinfo{pages}{869--873} (\bibinfo{year}{2021}).

\bibitem{he2022graphene}
\bibinfo{author}{He, P.} \emph{et~al.}
\newblock \bibinfo{title}{Graphene moir\'e superlattices with giant quantum nonlinearity of chiral {B}loch electrons}.
\newblock \emph{\bibinfo{journal}{Nat. Nanotechnol.}} \textbf{\bibinfo{volume}{17}}, \bibinfo{pages}{378--383} (\bibinfo{year}{2022}).

\bibitem{itahashi2022giant}
\bibinfo{author}{Itahashi, Y.~M.} \emph{et~al.}
\newblock \bibinfo{title}{Giant second harmonic transport under time-reversal symmetry in a trigonal superconductor}.
\newblock \emph{\bibinfo{journal}{Nat. Commun.}} \textbf{\bibinfo{volume}{13}}, \bibinfo{pages}{1659} (\bibinfo{year}{2022}).

\bibitem{zhang2022non}
\bibinfo{author}{Zhang, Z.} \emph{et~al.}
\newblock \bibinfo{title}{Non-reciprocal charge transport in an intrinsic magnetic topological insulator {M}n{B}i$_2${T}e$_4$}.
\newblock \emph{\bibinfo{journal}{Nat. Commun.}} \textbf{\bibinfo{volume}{13}}, \bibinfo{pages}{6191} (\bibinfo{year}{2022}).

\bibitem{ando2015topological}
\bibinfo{author}{Ando, Y.} \& \bibinfo{author}{Fu, L.}
\newblock \bibinfo{title}{Topological crystalline insulators and topological superconductors: from concepts to materials}.
\newblock \emph{\bibinfo{journal}{Annu. Rev. Condens. Matter Phys.}} \textbf{\bibinfo{volume}{6}}, \bibinfo{pages}{361--381} (\bibinfo{year}{2015}).

\bibitem{wu2022the}
\bibinfo{author}{Wu, H.} \emph{et~al.}
\newblock \bibinfo{title}{The field-free {J}osephson diode in a van der {W}aals heterostructure}.
\newblock \emph{\bibinfo{journal}{Nature}} \textbf{\bibinfo{volume}{604}}, \bibinfo{pages}{653--656} (\bibinfo{year}{2022}).

\bibitem{gao2023quantum}
\bibinfo{author}{Gao, A.} \emph{et~al.}
\newblock \bibinfo{title}{Quantum metric nonlinear {H}all effect in a topological antiferromagnetic heterostructure}.
\newblock \emph{\bibinfo{journal}{Science}} \textbf{\bibinfo{volume}{381}}, \bibinfo{pages}{181--186} (\bibinfo{year}{2023}).

\bibitem{wang2023quantum}
\bibinfo{author}{Wang, N.} \emph{et~al.}
\newblock \bibinfo{title}{Quantum-metric-induced nonlinear transport in a topological antiferromagnet}.
\newblock \emph{\bibinfo{journal}{Nature}} \textbf{\bibinfo{volume}{621}}, \bibinfo{pages}{487--492} (\bibinfo{year}{2023}).

\bibitem{zhang2020higher}
\bibinfo{author}{Zhang, C.-P.}, \bibinfo{author}{Gao, X.-J.}, \bibinfo{author}{Xie, Y.-M.}, \bibinfo{author}{Po, H.~C.} \& \bibinfo{author}{Law, K.~T.}
\newblock \bibinfo{title}{Higher-order nonlinear anomalous {H}all effects induced by {B}erry curvature multipoles.}
\newblock \emph{\bibinfo{journal}{Phys. Rev. B}} \textbf{\bibinfo{volume}{107}}, \bibinfo{pages}{115142} (\bibinfo{year}{2023}).

\bibitem{Wang2021Intrinsic}
\bibinfo{author}{Wang, C.}, \bibinfo{author}{Gao, Y.} \& \bibinfo{author}{Xiao, D.}
\newblock \bibinfo{title}{Intrinsic nonlinear {H}all effect in antiferromagnetic tetragonal {C}u{M}n{A}s}.
\newblock \emph{\bibinfo{journal}{Phys. Rev. Lett.}} \textbf{\bibinfo{volume}{127}}, \bibinfo{pages}{277201} (\bibinfo{year}{2021}).

\bibitem{liu2021intrinsic}
\bibinfo{author}{Liu, H.} \emph{et~al.}
\newblock \bibinfo{title}{Intrinsic second-order anomalous {H}all effect and its application in compensated antiferromagnets}.
\newblock \emph{\bibinfo{journal}{Phys. Rev. Lett.}} \textbf{\bibinfo{volume}{127}}, \bibinfo{pages}{277202} (\bibinfo{year}{2021}).

\bibitem{holder2021mixed}
\bibinfo{author}{Holder, T.}, \bibinfo{author}{Kaplan, D.}, \bibinfo{author}{Ilan, R.} \& \bibinfo{author}{Yan, B.}
\newblock \bibinfo{title}{Mixed axial-gravitational anomaly from emergent curved spacetime in nonlinear charge transport.} \bibinfo{pages}{Preprint at https://arxiv.org/abs/2111.07780} (\bibinfo{year}{2021}).

\bibitem{lahiri2022intrinsic}
\bibinfo{author}{Lahiri, S.}, \bibinfo{author}{Das, K.}, \bibinfo{author}{Culcer, D.} \& \bibinfo{author}{Agarwal, A.}
\newblock \bibinfo{title}{Intrinsic nonlinear conductivity induced by the quantum metric dipole.}
\newblock \emph{\bibinfo{journal}{Phys. Rev. B}} \textbf{\bibinfo{volume}{108}}, \bibinfo{pages}{L201405} (\bibinfo{year}{2023}).

\bibitem{Smith2022momentum}
\bibinfo{author}{Smith, T.~B.}, \bibinfo{author}{Pullasseri, L.} \& \bibinfo{author}{Srivastava, A.}
\newblock \bibinfo{title}{Momentum-space gravity from the quantum geometry and entropy of {B}loch electrons}.
\newblock \emph{\bibinfo{journal}{Phys. Rev. Res.}} \textbf{\bibinfo{volume}{4}}, \bibinfo{pages}{013217} (\bibinfo{year}{2022}).

\bibitem{ma2022anomalous}
\bibinfo{author}{Ma, D.}, \bibinfo{author}{Arora, A.}, \bibinfo{author}{Vignale, G.} \& \bibinfo{author}{Song, J.~C.}
\newblock \bibinfo{title}{Anomalous skew-scattering nonlinear {H}all effect and chiral photocurrents in $\mathcal{PT}$-symmetric antiferromagnets.}
\newblock \emph{\bibinfo{journal}{Phys. Rev. Lett.}} \textbf{\bibinfo{volume}{131}}, \bibinfo{pages}{076601} (\bibinfo{year}{2023}).

\bibitem{zhang2022efficient}
\bibinfo{author}{Zhang, P.} \emph{et~al.}
\newblock \bibinfo{title}{Control of {N}\'eel vector with spin-orbit torques in an antiferromagnetic insulator with tilted easy plane}.
\newblock \emph{\bibinfo{journal}{Phys. Rev. Lett.}} \textbf{\bibinfo{volume}{129}}, \bibinfo{pages}{017203} (\bibinfo{year}{2022}).

\bibitem{Zhang2022diodic}
\bibinfo{author}{Zhang, N.~J.} \emph{et~al.}
\newblock \bibinfo{title}{Angle-resolved transport non-reciprocity and spontaneous symmetry breaking in twisted trilayer graphene}.
\newblock \emph{\bibinfo{journal}{Nature Mater.}} \textbf{\bibinfo{volume}{23}}, \bibinfo{pages}{356--362} (\bibinfo{year}{2024}).

\bibitem{hayami2022nonlinear}
\bibinfo{author}{Hayami, S.} \& \bibinfo{author}{Yatsushiro, M.}
\newblock \bibinfo{title}{Nonlinear nonreciprocal transport in antiferromagnets free from spin-orbit coupling}.
\newblock \emph{\bibinfo{journal}{Phys. Rev. B}} \textbf{\bibinfo{volume}{106}}, \bibinfo{pages}{014420} (\bibinfo{year}{2022}).

\bibitem{kaplan2022unification}
\bibinfo{author}{Kaplan, D.}, \bibinfo{author}{Holder, T.} \& \bibinfo{author}{Yan, B.}
\newblock \bibinfo{title}{Unification of nonlinear anomalous {H}all effect and nonreciprocal magnetoresistance in metals by the quantum geometry}.
\newblock \emph{\bibinfo{journal}{Phys. Rev. Lett.}} \textbf{\bibinfo{volume}{132}}, \bibinfo{pages}{026301} (\bibinfo{year}{2024}).

\bibitem{huang2023nonlinear}
\bibinfo{author}{Huang, Y.-X.} \emph{et~al.}
\newblock \bibinfo{title}{Nonlinear current response of two-dimensional systems under in-plane magnetic field.}
\newblock \emph{\bibinfo{journal}{Phys. Rev. B}} \textbf{\bibinfo{volume}{108}}, \bibinfo{pages}{075155} (\bibinfo{year}{2023}).

\bibitem{huang2023Scaling}
\bibinfo{author}{Huang, Y.-X.}, \bibinfo{author}{Xiao, C.}, \bibinfo{author}{Yang, S.} \& \bibinfo{author}{Li, X.}
\newblock \bibinfo{title}{Scaling law for time-reversal-odd nonlinear transport.} \bibinfo{pages}{Preprint at https://arxiv.org/abs/2311.01219} (\bibinfo{year}{2023}).

\bibitem{atencia2023disorder}
\bibinfo{author}{Atencia, R.~B.}, \bibinfo{author}{Xiao, D.} \& \bibinfo{author}{Culcer, D.}
\newblock \bibinfo{title}{Disorder in the non-linear anomalous hall effect of $\mathcal{PT}$-symmetric {D}irac fermions.}
\newblock \emph{\bibinfo{journal}{Phys. Rev. B}} \textbf{\bibinfo{volume}{108}}, \bibinfo{pages}{L201115} (\bibinfo{year}{2023}).

\bibitem{kaplan2023general}
\bibinfo{author}{Kaplan, D.}, \bibinfo{author}{Holder, T.} \& \bibinfo{author}{Yan, B.}
\newblock \bibinfo{title}{General nonlinear {H}all current in magnetic insulators beyond the quantum anomalous {H}all effect}.
\newblock \emph{\bibinfo{journal}{Nature Commun.}} \textbf{\bibinfo{volume}{14}}, \bibinfo{pages}{3053} (\bibinfo{year}{2023}).

\bibitem{cheong2020seeing}
\bibinfo{author}{Cheong, S.-W.}, \bibinfo{author}{Fiebig, M.}, \bibinfo{author}{Wu, W.}, \bibinfo{author}{Chapon, L.} \& \bibinfo{author}{Kiryukhin, V.}
\newblock \bibinfo{title}{Seeing is believing: visualization of antiferromagnetic domains}.
\newblock \emph{\bibinfo{journal}{npj Quantum Mater.}} \textbf{\bibinfo{volume}{5}}, \bibinfo{pages}{1--10} (\bibinfo{year}{2020}).

\bibitem{Otrokov2019a}
\bibinfo{author}{Otrokov, M.~M.} \emph{et~al.}
\newblock \bibinfo{title}{Prediction and observation of an antiferromagnetic topological insulator}.
\newblock \emph{\bibinfo{journal}{Nature}} \textbf{\bibinfo{volume}{576}}, \bibinfo{pages}{416--422} (\bibinfo{year}{2019}).

\bibitem{Otrokov2019unique}
\bibinfo{author}{Otrokov, M.~M.} \emph{et~al.}
\newblock \bibinfo{title}{Unique thickness-dependent properties of the van der {W}aals interlayer antiferromagnet {M}n{B}i$_2${T}e$_4$ films}.
\newblock \emph{\bibinfo{journal}{Phys. Rev. Lett.}} \textbf{\bibinfo{volume}{122}}, \bibinfo{pages}{107202} (\bibinfo{year}{2019}).

\bibitem{Li2019a}
\bibinfo{author}{Li, J.} \emph{et~al.}
\newblock \bibinfo{title}{Intrinsic magnetic topological insulators in van der {W}aals layered {M}n{B}i$_2${T}e$_4$-family materials}.
\newblock \emph{\bibinfo{journal}{Sci. Adv.}} \textbf{\bibinfo{volume}{5}}, \bibinfo{pages}{eaaw5685} (\bibinfo{year}{2019}).

\bibitem{Zhang2019a}
\bibinfo{author}{Zhang, D.} \emph{et~al.}
\newblock \bibinfo{title}{Topological axion states in the magnetic insulator {M}n{B}i$_2${T}e$_4$ with the quantized magnetoelectric effect}.
\newblock \emph{\bibinfo{journal}{Phys. Rev. Lett.}} \textbf{\bibinfo{volume}{122}}, \bibinfo{pages}{206401} (\bibinfo{year}{2019}).

\bibitem{Hao2019gapless}
\bibinfo{author}{Hao, Y.-J.} \emph{et~al.}
\newblock \bibinfo{title}{Gapless surface dirac cone in antiferromagnetic topological insulator {M}n{B}i$_2${T}e$_4$}.
\newblock \emph{\bibinfo{journal}{Phys. Rev. X}} \textbf{\bibinfo{volume}{9}}, \bibinfo{pages}{041038} (\bibinfo{year}{2019}).

\bibitem{Zhang2020}
\bibinfo{author}{Zhang, R.-X.}, \bibinfo{author}{Wu, F.} \& \bibinfo{author}{Das~Sarma, S.}
\newblock \bibinfo{title}{M\"obius insulator and higher-order topology in {M}n{B}i$_{2n}${T}e$_{3n+1}$}.
\newblock \emph{\bibinfo{journal}{Phys. Rev. Lett.}} \textbf{\bibinfo{volume}{124}}, \bibinfo{pages}{136407} (\bibinfo{year}{2020}).

\bibitem{Liu2020}
\bibinfo{author}{Liu, Z.} \& \bibinfo{author}{Wang, J.}
\newblock \bibinfo{title}{Anisotropic topological magnetoelectric effect in axion insulators}.
\newblock \emph{\bibinfo{journal}{Phys. Rev. B}} \textbf{\bibinfo{volume}{101}}, \bibinfo{pages}{205130} (\bibinfo{year}{2020}).

\bibitem{Deng2020}
\bibinfo{author}{Deng, Y.} \emph{et~al.}
\newblock \bibinfo{title}{Quantum anomalous {H}all effect in intrinsic magnetic topological insulator {M}n{B}i$_2${T}e$_4$}.
\newblock \emph{\bibinfo{journal}{Science}} \textbf{\bibinfo{volume}{367}}, \bibinfo{pages}{895--900} (\bibinfo{year}{2020}).

\bibitem{Liu2020a}
\bibinfo{author}{Liu, C.} \emph{et~al.}
\newblock \bibinfo{title}{Robust axion insulator and {C}hern insulator phases in a two-dimensional antiferromagnetic topological insulator}.
\newblock \emph{\bibinfo{journal}{Nat. Mater.}} \textbf{\bibinfo{volume}{19}}, \bibinfo{pages}{522--527} (\bibinfo{year}{2020}).

\bibitem{Trang2020}
\bibinfo{author}{Trang, C.~X.} \emph{et~al.}
\newblock \bibinfo{title}{Crossover from 2{D} ferromagnetic insulator to wide bandgap quantum anomalous hall insulator in ultra-thin {M}n{B}i$_2${T}e$_4$}.
\newblock \emph{\bibinfo{journal}{ACS Nano}} \textbf{\bibinfo{volume}{15}}, \bibinfo{pages}{13444--13452} (\bibinfo{year}{2021}).

\bibitem{deng2021high}
\bibinfo{author}{Deng, H.} \emph{et~al.}
\newblock \bibinfo{title}{High-temperature quantum anomalous {H}all regime in a {MnBi$_2$Te$_4$/Bi$_2$Te$_3$} superlattice}.
\newblock \emph{\bibinfo{journal}{Nat. Phys.}} \textbf{\bibinfo{volume}{17}}, \bibinfo{pages}{36--42} (\bibinfo{year}{2021}).

\bibitem{yang2021odd}
\bibinfo{author}{Yang, S.} \emph{et~al.}
\newblock \bibinfo{title}{Odd-even layer-number effect and layer-dependent magnetic phase diagrams in {M}n{B}i$_2${T}e$_4$}.
\newblock \emph{\bibinfo{journal}{Phys. Rev. X}} \textbf{\bibinfo{volume}{11}}, \bibinfo{pages}{011003} (\bibinfo{year}{2021}).

\bibitem{Ovchinnikov2020}
\bibinfo{author}{Ovchinnikov, D.} \emph{et~al.}
\newblock \bibinfo{title}{Intertwined topological and magnetic orders in atomically thin {C}hern insulator {M}n{B}i$_2${T}e$_4$}.
\newblock \emph{\bibinfo{journal}{Nano Lett.}} \textbf{\bibinfo{volume}{21}}, \bibinfo{pages}{2544--2550} (\bibinfo{year}{2021}).

\bibitem{gao2021layer}
\bibinfo{author}{Gao, A.} \emph{et~al.}
\newblock \bibinfo{title}{Layer {H}all effect in a 2{D} topological axion antiferromagnet}.
\newblock \emph{\bibinfo{journal}{Nature}} \textbf{\bibinfo{volume}{595}}, \bibinfo{pages}{521--525} (\bibinfo{year}{2021}).

\bibitem{cai2021electric}
\bibinfo{author}{Cai, J.} \emph{et~al.}
\newblock \bibinfo{title}{Electric control of a canted-antiferromagnetic {C}hern insulator}.
\newblock \emph{\bibinfo{journal}{Nat. Commun.}} \textbf{\bibinfo{volume}{13}}, \bibinfo{pages}{1668} (\bibinfo{year}{2022}).

\bibitem{li2021nonlocal}
\bibinfo{author}{Li, Y.} \emph{et~al.}
\newblock \bibinfo{title}{Giant nonlocal edge conduction in the axion insulator state of {M}n{B}i$_2${T}e$_4$.}
\newblock \emph{\bibinfo{journal}{Sci. Bull.}} \textbf{\bibinfo{volume}{68}}, \bibinfo{pages}{1252--1258} (\bibinfo{year}{2023}).

\bibitem{tai2021polarity}
\bibinfo{author}{Tai, L.} \emph{et~al.}
\newblock \bibinfo{title}{Distinguishing two-component anomalous {H}all effect from topological {H}all effect in magnetic topological insulator {M}n{B}i$_2${T}e$_4$.}
\newblock \emph{\bibinfo{journal}{ACS Nano}} \textbf{\bibinfo{volume}{16}}, \bibinfo{pages}{17336--17346} (\bibinfo{year}{2022}).

\bibitem{Qiu2023axion}
\bibinfo{author}{Qiu, J.-X.} \emph{et~al.}
\newblock \bibinfo{title}{Axion optical induction of antiferromagnetic order}.
\newblock \emph{\bibinfo{journal}{Nat. Mater.}} \textbf{\bibinfo{volume}{22}}, \bibinfo{pages}{583--590} (\bibinfo{year}{2023}).

\bibitem{iyama2013magnetoelectric}
\bibinfo{author}{Iyama, A.} \& \bibinfo{author}{Kimura, T.}
\newblock \bibinfo{title}{Magnetoelectric hysteresis loops in {C}r$_2${O}$_3$ at room temperature}.
\newblock \emph{\bibinfo{journal}{Phys. Rev. B}} \textbf{\bibinfo{volume}{87}}, \bibinfo{pages}{180408} (\bibinfo{year}{2013}).

\bibitem{jiang2018electric}
\bibinfo{author}{Jiang, S.}, \bibinfo{author}{Shan, J.} \& \bibinfo{author}{Mak, K.~F.}
\newblock \bibinfo{title}{Electric-field switching of two-dimensional van der {W}aals magnets}.
\newblock \emph{\bibinfo{journal}{Nat. Mater.}} \textbf{\bibinfo{volume}{17}}, \bibinfo{pages}{406--410} (\bibinfo{year}{2018}).

\bibitem{Krichevtsov1993spontaneous}
\bibinfo{author}{Krichevtsov, B.~B.}, \bibinfo{author}{Pavlov, V.~V.}, \bibinfo{author}{Pisarev, R.~V.} \& \bibinfo{author}{Gridnev, V.~N.}
\newblock \bibinfo{title}{Spontaneous non-reciprocal reflection of light from antiferromagnetic {C}r$_2${O}$_3$}.
\newblock \emph{\bibinfo{journal}{J. Phys. Condens. Matter}} \textbf{\bibinfo{volume}{5}}, \bibinfo{pages}{8233} (\bibinfo{year}{1993}).

\bibitem{ahn2022theory}
\bibinfo{author}{Ahn, J.}, \bibinfo{author}{Xu, S.-Y.} \& \bibinfo{author}{Vishwanath, A.}
\newblock \bibinfo{title}{Theory of optical axion electrodynamics and application to the {K}err effect in topological antiferromagnets}.
\newblock \emph{\bibinfo{journal}{Nat. Commun.}} \textbf{\bibinfo{volume}{13}}, \bibinfo{pages}{7615} (\bibinfo{year}{2022}).

\bibitem{kovalev2010anomalous}
\bibinfo{author}{Kovalev, A.~A.}, \bibinfo{author}{Sinova, J.} \& \bibinfo{author}{Tserkovnyak, Y.}
\newblock \bibinfo{title}{Anomalous {H}all effect in disordered multiband metals}.
\newblock \emph{\bibinfo{journal}{Phys. Rev. Lett.}} \textbf{\bibinfo{volume}{105}}, \bibinfo{pages}{036601} (\bibinfo{year}{2010}).

\bibitem{fujimoto2023observation}
\bibinfo{author}{Fujimoto, T.} \emph{et~al.}
\newblock \bibinfo{title}{{Observation of Terahertz Spin Hall Conductivity Spectrum in GaAs with Optical Spin Injection}.}
\newblock \emph{\bibinfo{journal}{Phys. Rev. Lett.}} \textbf{\bibinfo{volume}{132}}, \bibinfo{pages}{016301} (\bibinfo{year}{2023}).

\bibitem{walmsley2017determination}
\bibinfo{author}{Walmsley, P.} \& \bibinfo{author}{Fisher, I.}
\newblock \bibinfo{title}{Determination of the resistivity anisotropy of orthorhombic materials via transverse resistivity measurements}.
\newblock \emph{\bibinfo{journal}{Rev. Sci. Ins.}} \textbf{\bibinfo{volume}{88}}, \bibinfo{pages}{043901} (\bibinfo{year}{2017}).

\bibitem{orenstein2011optical}
\bibinfo{author}{Orenstein, J.}
\newblock \bibinfo{title}{Optical nonreciprocity in magnetic structures related to high-${T}_{\textrm{c}}$ superconductors}.
\newblock \emph{\bibinfo{journal}{Phys. Rev. Lett.}} \textbf{\bibinfo{volume}{107}}, \bibinfo{pages}{067002} (\bibinfo{year}{2011}).

\bibitem{varma2014gyrotropic}
\bibinfo{author}{Varma, C.}
\newblock \bibinfo{title}{Gyrotropic birefringence in the underdoped cuprates}.
\newblock \emph{\bibinfo{journal}{Europhys. Lett.}} \textbf{\bibinfo{volume}{106}}, \bibinfo{pages}{27001} (\bibinfo{year}{2014}).

\bibitem{beidenkopf2011spatial}
\bibinfo{author}{Beidenkopf, H.} \emph{et~al.}
\newblock \bibinfo{title}{Spatial fluctuations of helical {D}irac fermions on the surface of topological insulators}.
\newblock \emph{\bibinfo{journal}{Nature Phys.}} \textbf{\bibinfo{volume}{7}}, \bibinfo{pages}{939--943} (\bibinfo{year}{2011}).

\end{thebibliography}


\begin{thebibliography}{99}  
 \makeatletter
 \addtocounter{NAT@ctr}{79}
 \makeatother

\bibitem{Yan2019}
\bibinfo{author}{Yan, J.-Q.} \emph{et~al.}
\newblock \bibinfo{title}{Crystal growth and magnetic structure of {M}n{B}i$_2${T}e$_4$}.
\newblock \emph{\bibinfo{journal}{Phys. Rev. Mater.}}
  \textbf{\bibinfo{volume}{3}}, \bibinfo{pages}{064202} (\bibinfo{year}{2019}).
  \bibitem{Thiel2019}
\bibinfo{author}{Thiel, L.} \emph{et~al.}
\newblock \bibinfo{title}{Probing magnetism in 2{D} materials at the nanoscale
  with single-spin microscopy}.
\newblock \emph{\bibinfo{journal}{Science}} \textbf{\bibinfo{volume}{364}},
  \bibinfo{pages}{973--976} (\bibinfo{year}{2019}).
\bibitem{hsieh2011}
\bibinfo{author}{Hsieh, D.} \emph{et~al.}
\newblock \bibinfo{title}{Selective probing of photoinduced charge and spin
  dynamics in the bulk and surface of a topological insulator}.
\newblock \emph{\bibinfo{journal}{Phys. Rev. Lett.}}
  \textbf{\bibinfo{volume}{107}}, \bibinfo{pages}{077401}
  (\bibinfo{year}{2011}).
\bibitem{Vool2021}
\bibinfo{author}{Vool, U.} \emph{et~al.}
\newblock \bibinfo{title}{Imaging phonon-mediated hydrodynamic flow in {WT}e$_2$}.
\newblock \emph{\bibinfo{journal}{Nat. Phys.}}
  \textbf{\bibinfo{volume}{17}}, \bibinfo{pages}{1216--1220}
  (\bibinfo{year}{2021}).
\bibitem{Perdew1996}
\bibinfo{author}{Perdew, J. P.} \emph{et~al.}
\newblock \bibinfo{title}{Generalized Gradient Approximation Made Simple}.
\newblock \emph{\bibinfo{journal}{Phys. Rev. Lett.}}
  \textbf{\bibinfo{volume}{77}}, \bibinfo{pages}{3865--3868}
  (\bibinfo{year}{1996}).
\bibitem{Pizzi2020}
\bibinfo{author}{Pizzi, G.} \emph{et~al.}
\newblock \bibinfo{title}{Wannier90 as a community code: new features and applications}.
\newblock \emph{\bibinfo{journal}{J. Phys. Condens. Matter}}
  \textbf{\bibinfo{volume}{32}}, \bibinfo{pages}{165902}
  (\bibinfo{year}{2020}).
  
  \bibitem{steiner2016}
S.~Steiner, S.~Khmelevskyi, M.~Marsmann, G.~Kresse, Calculation of the magnetic
  anisotropy with projected-augmented-wave methodology and the case study of
  disordered {F}e$_{1- x}${C}o$_x$ alloys, {\it Phys. Rev. B\/} {\bf 93},
  224425 (2016).

  
  \end{thebibliography}
\vspace{0.5cm}
\vspace{0.3cm}
\textbf{Methods}
\\
\textbf{Bulk crystal growth}\\
Our MnBi$_2$Te$_4$ bulk crystals were grown by the Bi$_2$Te$_3$ flux method \cite{Yan2019}. Elemental Mn, Bi, and Te were mixed at a molar ratio of $15:170:270$, loaded in a crucible, and sealed in a quartz tube under one-third atmospheric pressure of Ar. The ampule was first heated to $900^{\circ}$C for $5$ hours. It was then moved to another furnace where it slowly cooled from $597^{\circ}$C to $587^{\circ}$C and stayed at $587^{\circ}$C for one day. Finally, MnBi$_2$Te$_4$ were obtained by centrifuging the ampule to separate the crystals from Bi$_2$Te$_3$ flux.
\vspace{3mm}
\\
\textbf{Sample fabrication}\\
To address the sensitive chemical nature of MnBi$_2$Te$_4$ flakes, all fabrication processes were completed in an argon environment without exposure to air, chemicals, or heat. Specifically, the argon-filled glovebox maintained O$_2$ and H$_2$O level below $0.01$ ppm and a dew point below $-96^{\circ}$C. The glovebox was attached to an e-beam evaporator, allowing us to make metal deposition without exposure to air. MnBi$_2$Te$_4$ was mechanically exfoliated onto a baked $300$-nm SiO$_2/$Si wafer using Scotch tape. MnBi$_2$Te$_4$ flakes with straight edges were deliberately chosen (those straight edges are likely to be along the crystalline direction). After scratching the flake to a rectangular/circular shape with a tip, a stencil mask technique \cite{gao2021layer} was used to make Cr/Au contacts on top of MnBi$_2$Te$_4$ without exposure to air or chemical. BN flakes were directly exfoliated onto a polydimethylsiloxane (PDMS) film, and a 10-30 nm thick BN flake was identified and transferred onto the MnBi$_2$Te$_4$ as the top gate dielectric layer. Next, a metal gate was evaporated onto the BN/MnBi$_2$Te$_4$ heterostructure. The contacts were aligned along the straight edges of MnBi$_2$Te$_4$. To check the angle between contact and MnBi$_2$Te$_4$ crystalline direction, the optical second harmonic generation (SHG) measurements were performed onto the sample after the transport measurements were done (BN and top gate were removed by blue tape before SHG measurements). This approach ensured MnBi$_2$Te$_4$ flakes were fresh for transport measurements.
\vspace{3mm}
\\
\textbf{Optical second harmonic generation measurements}\\
All optical SHG experiments were performed using a near-infrared femtosecond laser at room temperature. The light source is an amplified Yb:KGW laser (Pharos, LightConversion) emitting $168$ fs pulses at $1.2$ eV with a pulse energy of $100$ $\mu$J and a default repetition rate of $100$ kHz. All measurements were performed at normal incidence. The polarization of the incident laser was controlled using an achromatic half-wave plate, while a Glan-Laser (GL) polarizer prism was used as an analyzer to select the polarization of the outgoing SHG signal. Both half-wave plate and GL prism were mounted on motorized rotation stages.
At room temperature, the interior of thin MnBi$_2$Te$_4$ flake's crystal structure is centrosymmetric (group $D_{3d}$). Therefore, the SHG signals are only expected to originate from the surface (surface group $C_{3v}$). These results are consistent with previous SHG results on Bi$_2$Se$_3$ (same symmetry as MnBi$_2$Te$_4$ above $T_{\textrm{N}}$) \cite{hsieh2011}. 
\vspace{3mm}
\\
\textbf{Optical circular dichroism measurements}\\
Optical CD measurements were performed in the closed-loop magneto-optical cryostat OptiCool by Quantum Design (base temperature $\sim$2 K and B field $\pm$7 T) using a supercontinuum laser SuperK Extreme by NKT photonics (wavelength 500nm to 2500 nm, pulse width  12 ps at 1064 nm). A spectrometer SpectraPro-300i by Acton Research was used to select the wavelength. The wavelength used in the CD measurement is 946 nm. The beam went through a photoelastic modulator (PEM) operating at $\lambda/4$ retardation with a frequency of 50 kHz. After an optical chopper (1000 Hz), the beam was focused onto the sample by a super apochromatic objective (NA=0.42, Mitutoyo). The reflected beam went through the cryostat's top window, was collimated by the same objective, and was collected by an avalanche photodiode (APD). The corresponding reflection APD signal was analyzed by two lock-in amplifiers at 50 kHz (the PEM frequency) and 1000 Hz (the chopper frequency), respectively. The CD was the ratio of the 50 kHz and 1000 Hz signals. Spatial imaging was achieved using a galvo scanning mirror system.
\vspace{3mm}
\\
\textbf{Nitrogen vacancy measurements}\\
The magnetization of 6SL and 7SL MnBi$_2$Te$_4$ flakes was measured with a home-built scanning NV magnetometry microscope, at base temperature of 4 K and in the He$_4$ vapor environment. Details of the setup is described in Ref.~\cite{Vool2021}. The flake is field-cooled and measured with an external field aligned with the NV axis (out-of-plane field of 80 mT, and in-plane field of 140 mT). The NV center measures the static magnetic field generated by the odd-layered ferromagnetic flake at a stand-off height of about 100 nm, and the magnetization is reconstructed with the Fourier transform method similar to those described in Ref.~\cite{Thiel2019}. 
\vspace{3mm}
\\
\textbf{Nonlinear electrical transport measurements}\\
Electrical transport measurements were carried out in a PPMS (Quantum Design DynaCool). The base temperature is 1.65 K and maximum magnetic field is 9 T. The gate voltages were applied by Keithley 2400 source meters. Longitudinal and transverse voltages were measured simultaneously. Both first- and second-harmonic signals were collected by standard Lock-in techniques (Stanford Research Systems Model SR830) with excitation frequencies between 1-600 Hz. 
\vspace{3mm}
\\
\textbf{Sum frequency generation measurements}\\
In the SFG measurements, we injected two currents with frequencies $\omega_1$ and $\omega_2$, and detected the SFG voltage $V^{\omega_1+\omega_2}$. By separating the two currents in frequency domain, it is easier to control their directions separately. The SFG voltage can be expressed as $V^{\omega_1+\omega_2}\propto \sin(\omega_1+\omega_2)t$. The standard lock-in detectors (SR830) cannot directly lock to the $\omega_1+\omega_2$ frequency. Instead, we measured $\sin\omega_1t\sin\omega_2t$, which directly relates to $\sin(\omega_1+\omega_2)t$ by the angle addition theorem. We chose $\omega_1=547.1$ Hz and $\omega_2=1.77$ Hz, so that $\omega_1\gg\omega_2$. In order to properly define the current trajectory of the $\omega_1$ current and $\omega_2$ current, a 40 H inductor and 1 $\mu$F capacitor were used. To probe $\sin\omega_1t\sin\omega_2t$, the SFG voltage signal was fed to the first lock-in (lock-in A), which was locked to $\omega_1$ and its integral time was set to $\frac{2\pi}{w_1}<t_1\ll\frac{2\pi}{w_2}$. The output was then fed to the second lock-in (lock-in B), which was locked to $\omega_2$ and its integral time was set to $t_2>\frac{2\pi}{w_2}$. The output of the second lock-in was $\sin\omega_1t\sin\omega_2t$. The measurement setups for $\sigma^{2\omega}_{yxx}$ and $\sigma^{2\omega}_{xyx}$ are shown in Extended Data Fig.~\ref{SFG}.
\vspace{3mm}
\\
\textbf{In-plane field effect measurements}\\
The in-plane field effect experiments are similar to SFG measurements except $\omega_2 \rightarrow 0$. The DC signal was applied by a Keithley 2400 source-meter along $y$ and the $\omega_1$ signal was applied by a lock-in along $x$. In order to properly define the trajectory of DC current and $\omega_1$ current, we connected a 40 H inductor in series in the DC current loop before grounding, and a 1 $\mu$F capacitor in series in the $\omega_1$ current loop before grounding. The in-plane field effect voltage $V_x^{\omega_1}$ can be expressed as $V_x^{\omega_1} = R_{xyx} I_x^{\omega_1} I_y^{\textrm{DC}}=(R_{xyx} I_y^{\textrm{DC}}) I_x^{\omega_1}=\Delta R_{xx} I_x^{\omega_1}$, where $\Delta R_{xx} = R_{xyx} I_y^{\textrm{DC}}$. The chiral inner product is small in the near DC transport regime. The $V_x^{\omega_1}$ was directly recorded by a standard lock-in detectors. The in-plane field effect resistance $\Delta R_{xx}$ was calculated by $\Delta R_{xx}=V_x^{\omega_1}/I_x^{\omega_1}$, which can be modulated by $I_y^{\textrm{DC}}$ (Fig.~\ref{Fig4}\textbf{d}).
\vspace{3mm}
\\
\textbf{Wireless radio frequency (RF) rectification measurements}\\
We set up a simple experiment to harvest wireless RF signals and recorded the generated DC signals. The RF signal generator was Hittite HMC-T2220. The samples were connected to a low-temperature probe. The RF signals went into the low-temperature probe through a coaxial cable and the other end of the coaxial cable was connected to an antenna. The antenna was made by a 50-mm long conducting wire with a $\sim$0.2-mm diameter. The end of the antenna was parallel to the sample and the spacing was $\sim$10 mm. The electrical field direction and power of the RF signals shone on the sample were not well defined. The DC voltage signals were recorded by Agilent 34401A Digital Multimeters.
\vspace{3mm}
\\
\textbf{The density functional theory calculations}\\
The density functional theory (DFT) based $ab$ initio calculations were performed using the projector augmented wave (PAW) pseudopotentials and a plane wave basis set as implemented in the VASP package. The spin-orbit coupling (SOC) was included following the standard implementation in the VASP package \cite{steiner2016}. A $9\times9\times1$ $\Gamma$-centered $k$-grid was used for the Brillouin zone integration. The kinetic energy cutoff for the plane wave basis was set to 400 eV. The generalized gradient approximation (GGA+U) scheme with an onsite Coulomb potential ($U = 5.0$ eV) was employed to treat the localized Mn  $3d$ orbitals~\cite{Perdew1996}. The Wannier function-based tight binding model was built using the Wannier90 code~\cite{Pizzi2020}.

%\bibliographystyle{naturemag}
%\bibliography{Topological_and_2D_02062022}

\vspace{0.5cm}

\textbf{Data availability:} The data that support the plots within this paper and other findings of this study are available from the corresponding author upon reasonable request.
\textbf{Code availability:} The computer code used in this study is available from the corresponding authors upon reasonable request.

\textbf{Acknowledgements:}  We thank Yang Gao, Cong Xiao Amit Agarwal, Junyeong Ahn, Shengyuan Yang, Fernando de Juan and Inti Sodemann for helpful discussions. Work in the SYX group was partly supported through the Center for the Advancement of Topological Semimetals (CATS), an Energy Frontier Research Center (EFRC) funded by the U.S. Department of Energy (DOE) Office of Science (fabrication and measurements), through the Ames National Laboratory under contract DE-AC-0207CH11358. and partly through Air Force Office of Scientific Research (AFOSR) grant FA9550-23-1-0040 (data analysis and manuscript writing). SYX acknowledges the Corning Fund for Faculty Development and support from the Alfred P. Sloan Foundation. SYX and DB were supported by the NSF Career DMR-2143177. CT and ZS acknowledge support from the Swiss National Science Foundation under project P2EZP2 191801 and P500PT 206914, respectively. YFL, SYX, DCB, YO, and LF were supported by the STC Center for Integrated Quantum Materials (CIQM), NSF Grant no. DMR-1231319. This work was performed in part at the Center for Nanoscale Systems (CNS) Harvard University, a member of the National Nanotechnology Coordinated Infrastructure Network (NNCI), which is supported by the National Science Foundation under NSF awards no. ECCS-2025158. Bulk single crystal growth and characterization of MnBi$_2$Te$_4$ were performed at UCLA, which were supported by the DOE, Office of Science, under Award Number DE-SC0021117. The work at Northeastern University was supported by the National Science Foundation through NSF-ExpandQISE award no. 2329067 and it benefited from the resources of Northeastern University's Advanced Scientific Computation Center and the Discovery Cluster, and the Quantum Materials and Sensing Institute. The work in the QM group was supported through the CATS, an EFRC funded by the DOE Office of Science, through the Ames National Laboratory under contract DE-AC02-07CH11358 (sample fabrication) and was partly supported by the Air Force Office of Scientific Research under Grant No. FA9550-22-1-0270 (manuscript writing) and the Alfred P. Sloan Foundation. LF and QM also acknowledge support from the NSF Convergence program (NSF ITE-2345084) and the Canadian Institute for Advanced Research. TRC was supported by the 2030 Cross-Generation Young Scholars Program from the National Science and Technology Council (NSTC) in Taiwan (Program no. MOST111-2628-M-006-003-MY3), National Cheng Kung University (NCKU), Taiwan, and National Center for Theoretical Sciences, Taiwan. TRC thanks to National Center for High-performance Computing (NCHC) of National Applied Research Laboratories (NARLabs) in Taiwan for providing computational and storage sources. This research was supported, in part, by Higher Education Sprout Project Ministry of Education to the Headquarters of University Advancement at NCKU. HL acknowledges the support of the National Science and Technology Council (NSTC) in Taiwan under grant number MOST 111-2112-M-001-057-MY3. The work at TIFR Mumbai was supported by the Department of Atomic Energy of the Government of India under Project no. 12-R\&D-TFR-5.10-0100 and benefited from the computational resources of TIFR Mumbai. KW and TT acknowledge support from the JSPS KAKENHI (Grant Numbers 20H00354, 21H05233, and 23H02052) and World Premier International Research Center Initiative (WPI), MEXT, Japan. SWC acknowledges partial support from the Harvard Quantum Initiative in Science and Engineering.

\textbf{Author Contributions Statement:} SYX conceived and supervised the project. AG fabricated the devices and performed the transport measurements with help from YFL, JXQ, DB, TD, HCL, CT, ZS, SCH,  TH, \color{black} DCB and QM. AG, SWC and SP performed the microwave rectification and nitrogen vacancy experiments with help from AY. AG and JXQ performed the optical CD measurement. AG performed the SHG measurements with help from CT, HCL, JXQ and YFL. CH, TQ and NN grew the bulk MnBi$_2$Te$_4$ single crystals.  BG made the theoretical studies including first-principles calculations with the help from YO, under the guidance of BS, AB, HL, LF, TRC. KW and TT grew the bulk h-BN single crystals. SYX, AG and QM wrote the manuscript with input from all authors.

\textbf{Competing financial interests:} The authors declare no competing financial interests.

\clearpage

\clearpage
%%%
%%
\clearpage

\setcounter{figure}{0}
\renewcommand{\figurename}{\textbf{Extended Data Fig.}}

\clearpage

\begin{figure*}[h]
\includegraphics[width=12cm]{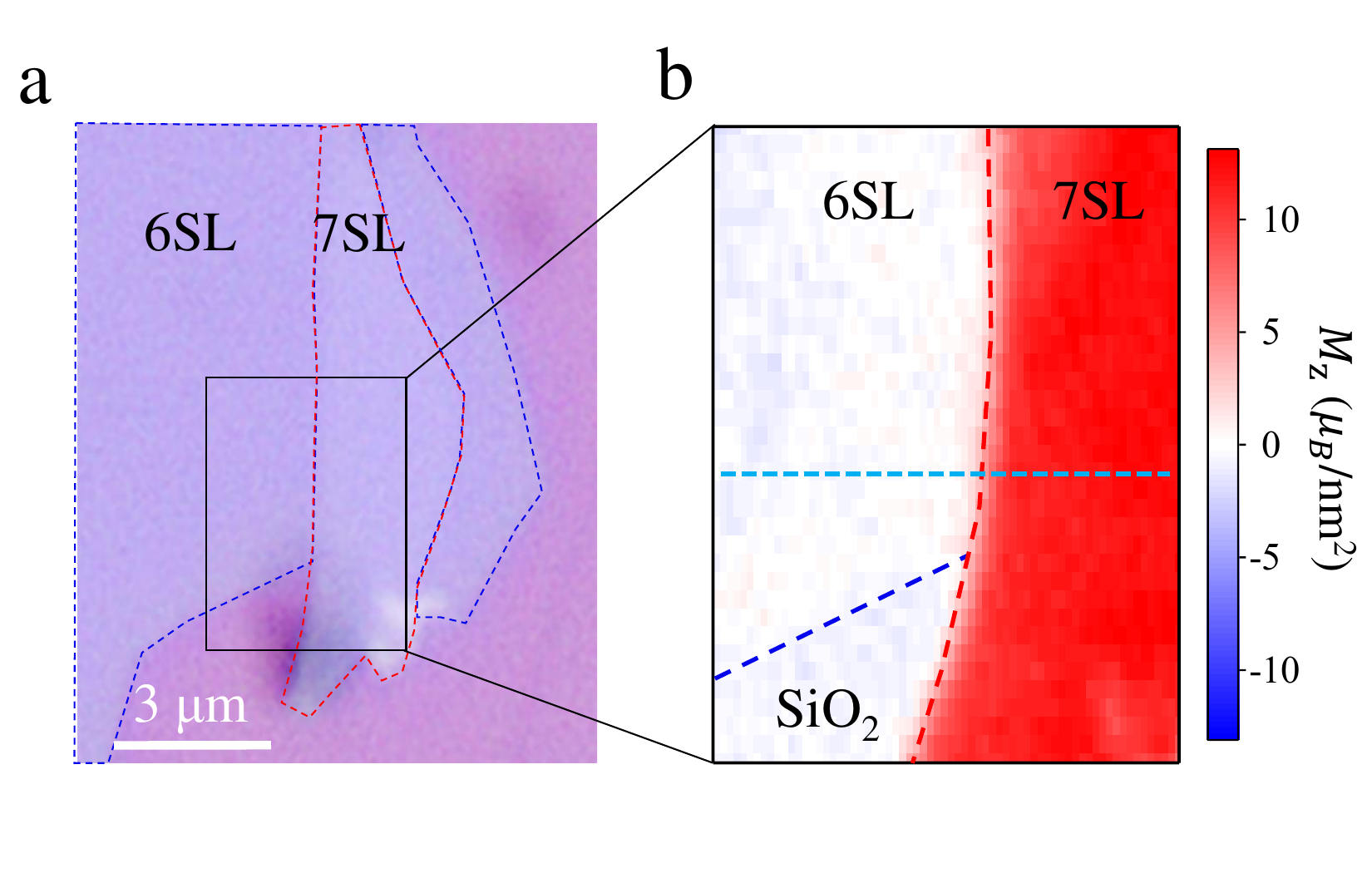}
\vspace{-0.5mm}
\caption{{\bf The fully compensated antiferromagnetism for 6SL MnBi$_2$Te$_4$.} \textbf{a,b}, Nitrogen vacancy (NV) center magnetometry measurement for even- and odd-layered MnBi$_2$Te$_4$ (outlined by blue and red dashed lines in the optical image of the sample in (a)). The magnetization is only observed in odd-layered MnBi$_2$Te$_4$ and negligible in even-layered MnBi$_2$Te$_4$. The cyan dotted line denotes the line cut data shown in Fig.~\ref{Fig1}\textbf{e}.}
\label{Compensated_AFM}
\end{figure*}

\clearpage
\begin{figure*}[h]
\centering
\includegraphics[width=12cm]{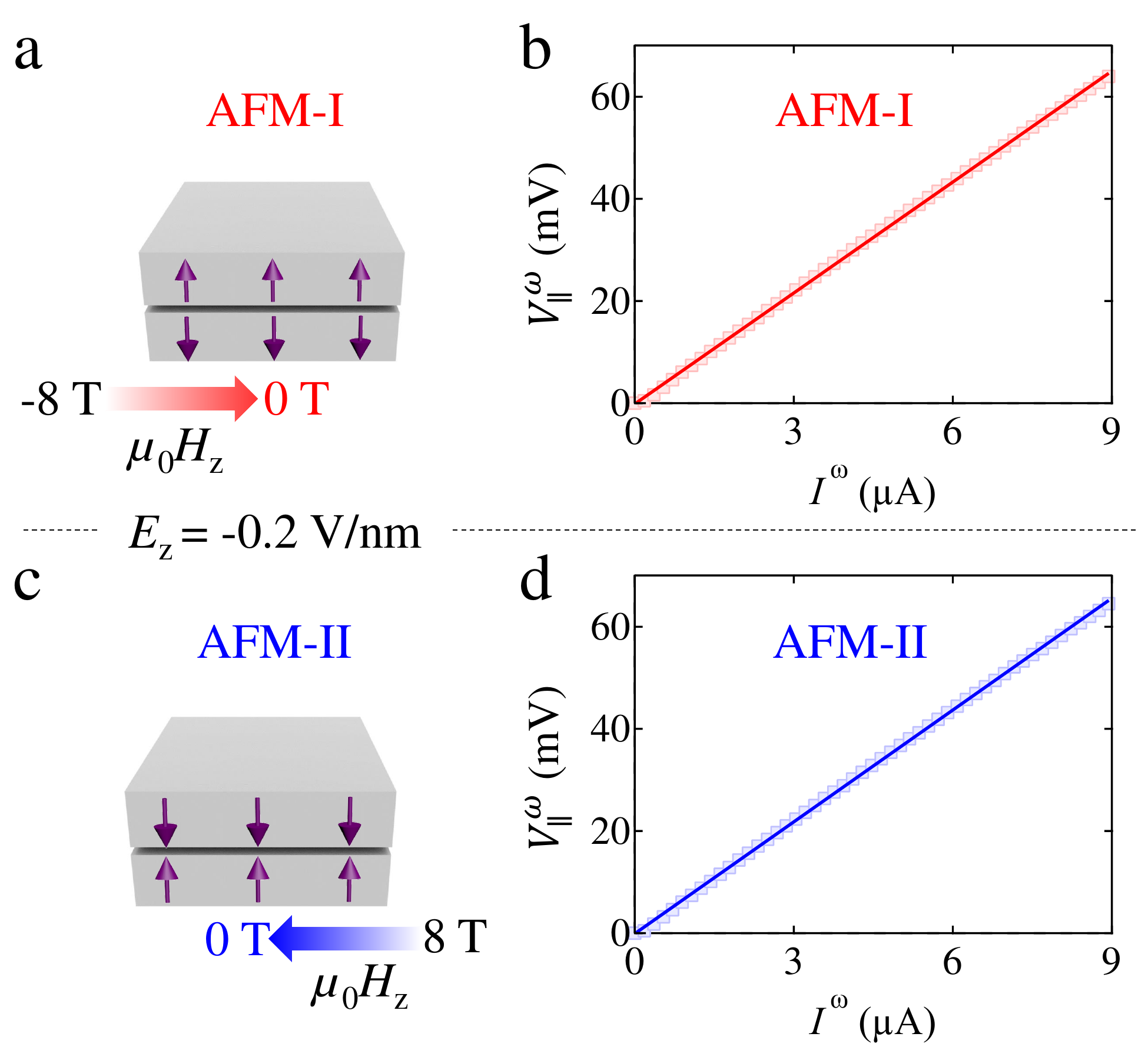}
\caption{\textbf{The linear and nonlinear signals for opposite AFM states in a 6SL MnBi$_2$Te$_4$.} \textbf{a,c}, At a finite $E_z$ field, AFM-I state can be prepared by sweeping $B_z$ from $-8$ T to $0$ T (panel a); AFM-II state can be prepared by sweeping $B_z$ from $+8$ T to $0$ T (panel c). \textbf{b}, The linear longitudinal voltage $V_\parallel^{\omega}$ as a function of injection current for AFM-I. \textbf{d} Same as (b) but for AFM-II. For different AFM states, the linear signal is the same. These data were taken from Device A and the current was along 30$^{\circ}$ in Fig.~\ref{Fig2}\textbf{g}.
}
\label{Linear_AFM}
\end{figure*}
\vspace{1cm} 

\clearpage
\begin{figure*}[h]
\includegraphics[width=16cm]{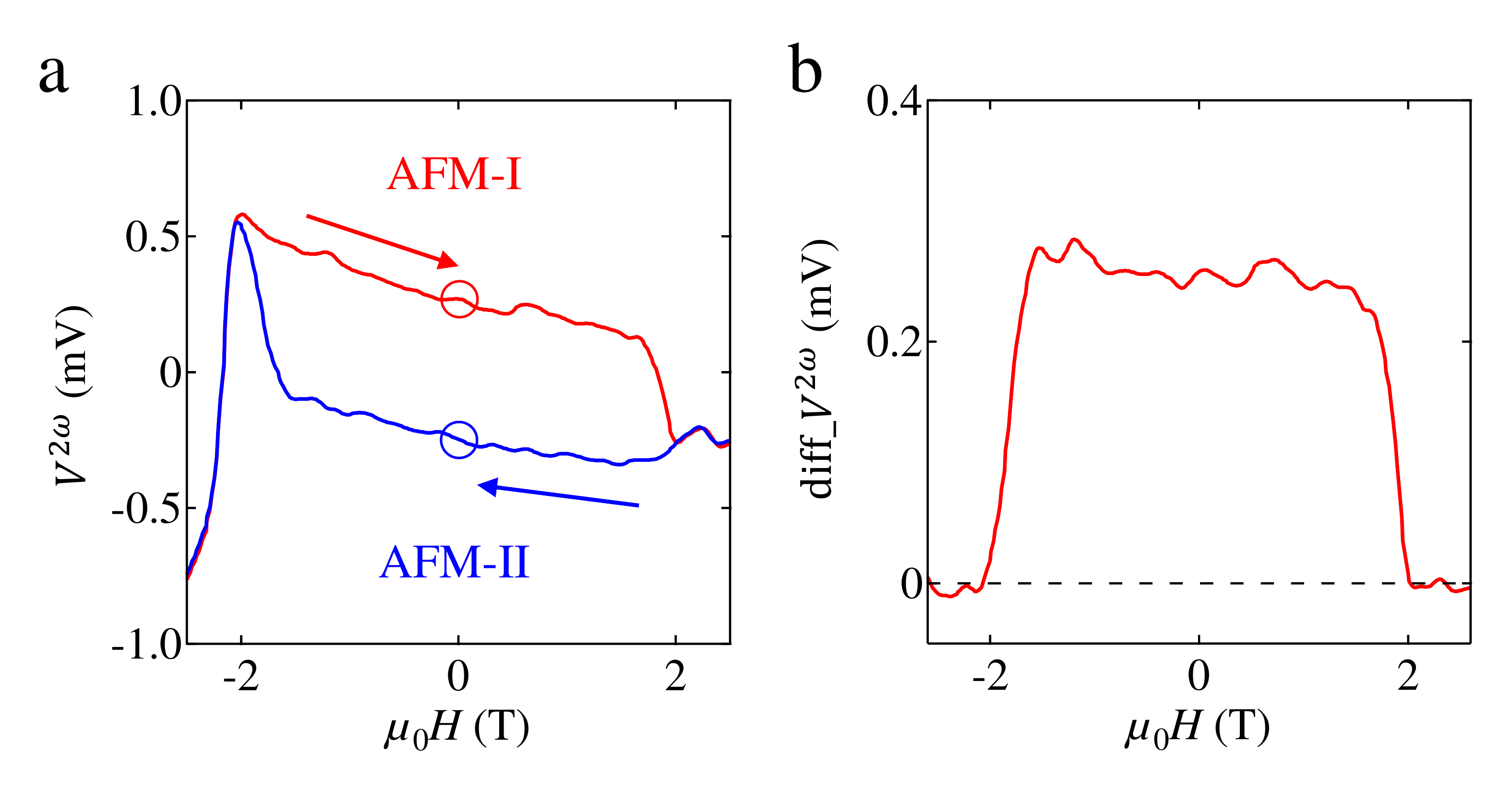}
\caption{\textbf{AFM order determined nonlinear signal.} \textbf{a,} Nonlinear voltage $V^{2\omega}$ as a function of external magnetic field $\mu_0H$. $V^{2\omega}$ is opposite for two AFM orders ($\vert\mu_0H\vert<$ 2 T). \textbf{b,} The nonlinear voltage difference ($\text{diff}\_V^{2\omega}=\frac{1}{2}(V^{2\omega}_{\text{AFM}-\textrm{I}}-V^{2\omega}_{\text{AFM}-\textrm{II}})$) between two AFM orders as a function of $\mu_0H$. }
\label{hysB_Nonlinear}
\end{figure*}

\clearpage
\begin{figure*}[h]
\includegraphics[width=15cm]{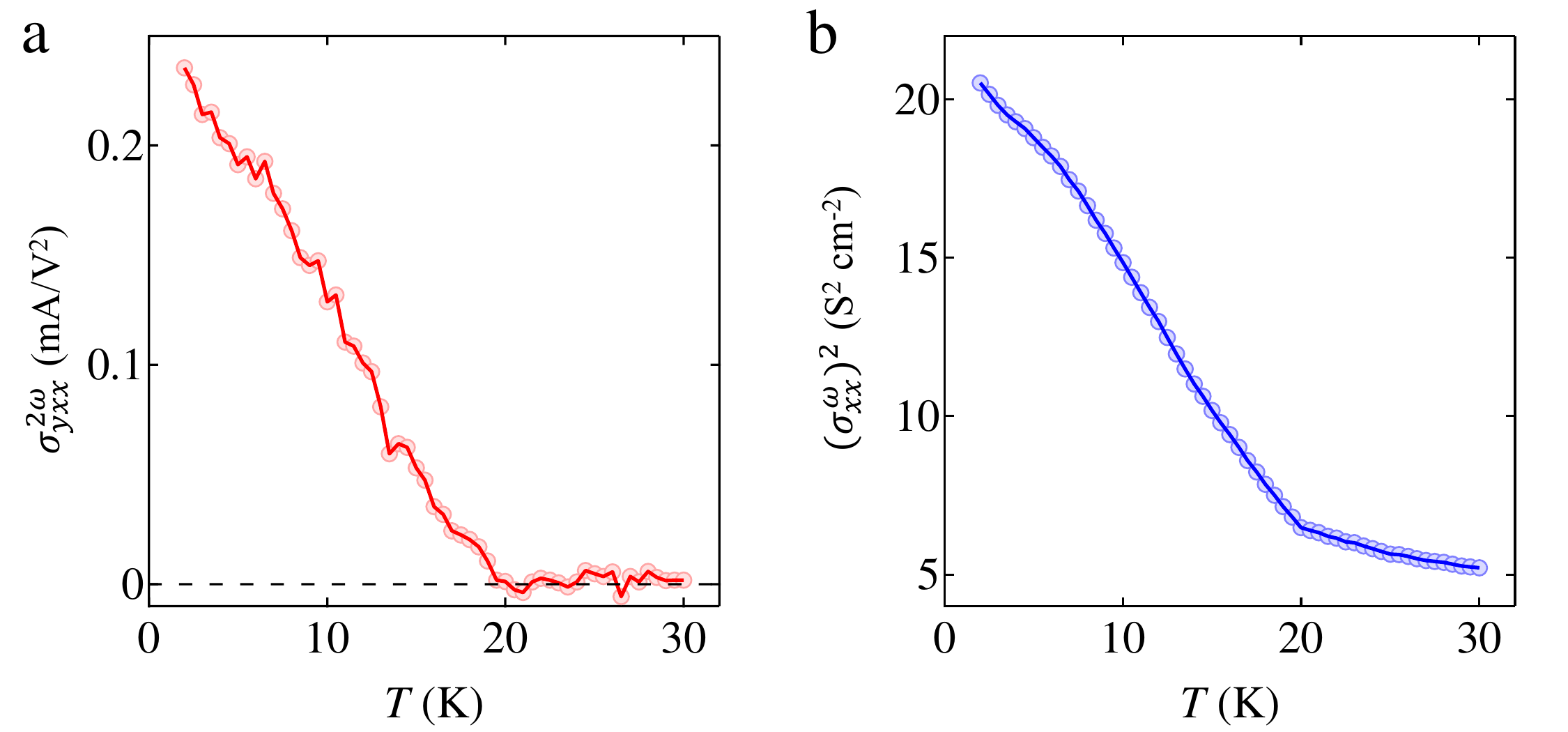}
\caption{\textbf{The temperature dependence of the nonlinear signal.} \textbf{a,b,} Nonlinear conductivity $\sigma_{yxx}^{2\omega}$ and square of linear conductivity $(\sigma_{xx}^{\omega})^2$ as a function of temperature. $\sigma_{yxx}^{2\omega}$ and $(\sigma_{xx}^{\omega})^2$ show a similar temperature dependence.}
\label{Scaling}
\end{figure*}

\clearpage
\begin{figure*}[h]
\includegraphics[width=15cm]{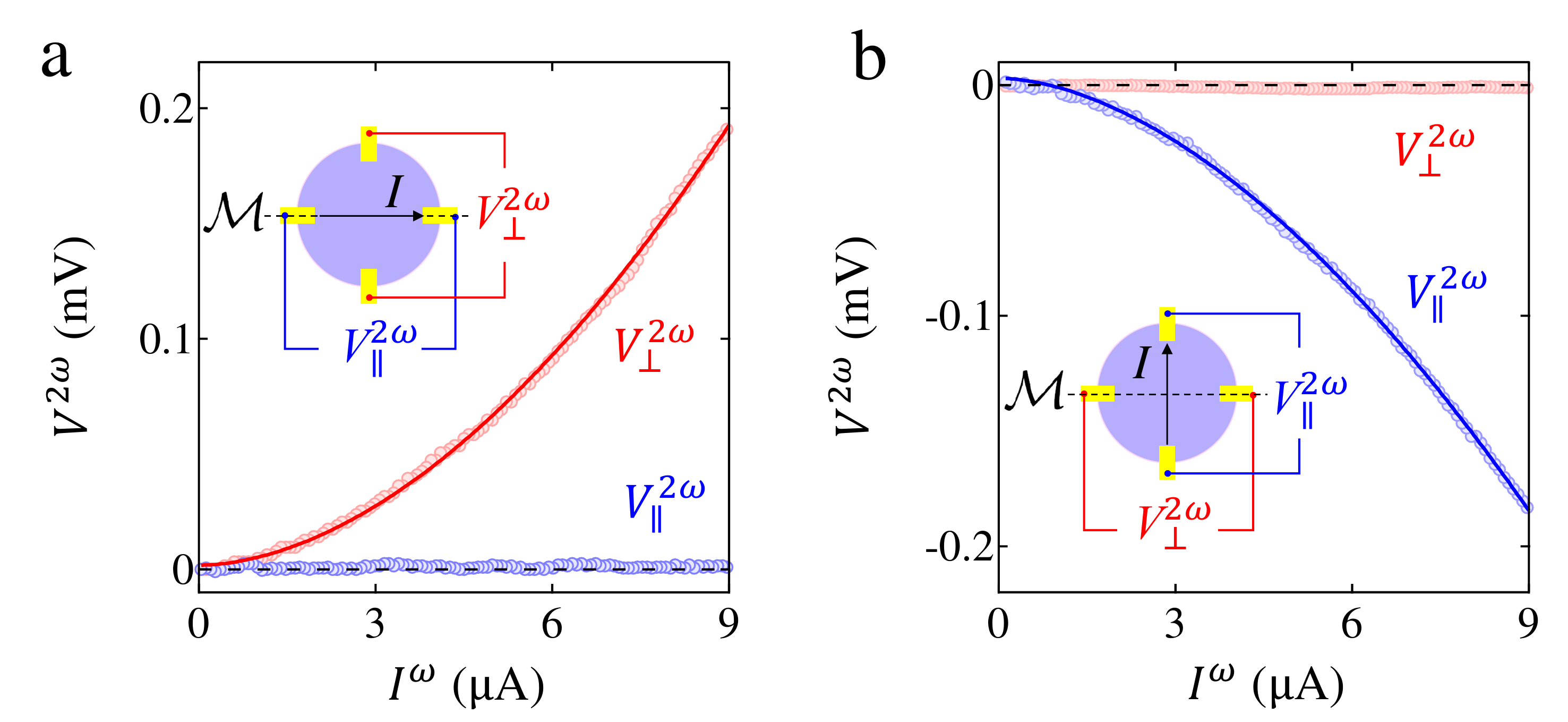}
\caption{\textbf{Nonlinear voltage versus current direction.} \textbf{a,b,} Longitudinal and transverse nonlinear voltage when the current is injected along $x$ (panel a) and along $y$ (panel b). A crystalline axes is along $x$ in Device A (Fig.~\ref{Fig2}\textbf{g,h}).}
\label{NLV_Direction}
\end{figure*}

\clearpage
\begin{figure*}[h]
\includegraphics[width=17cm]{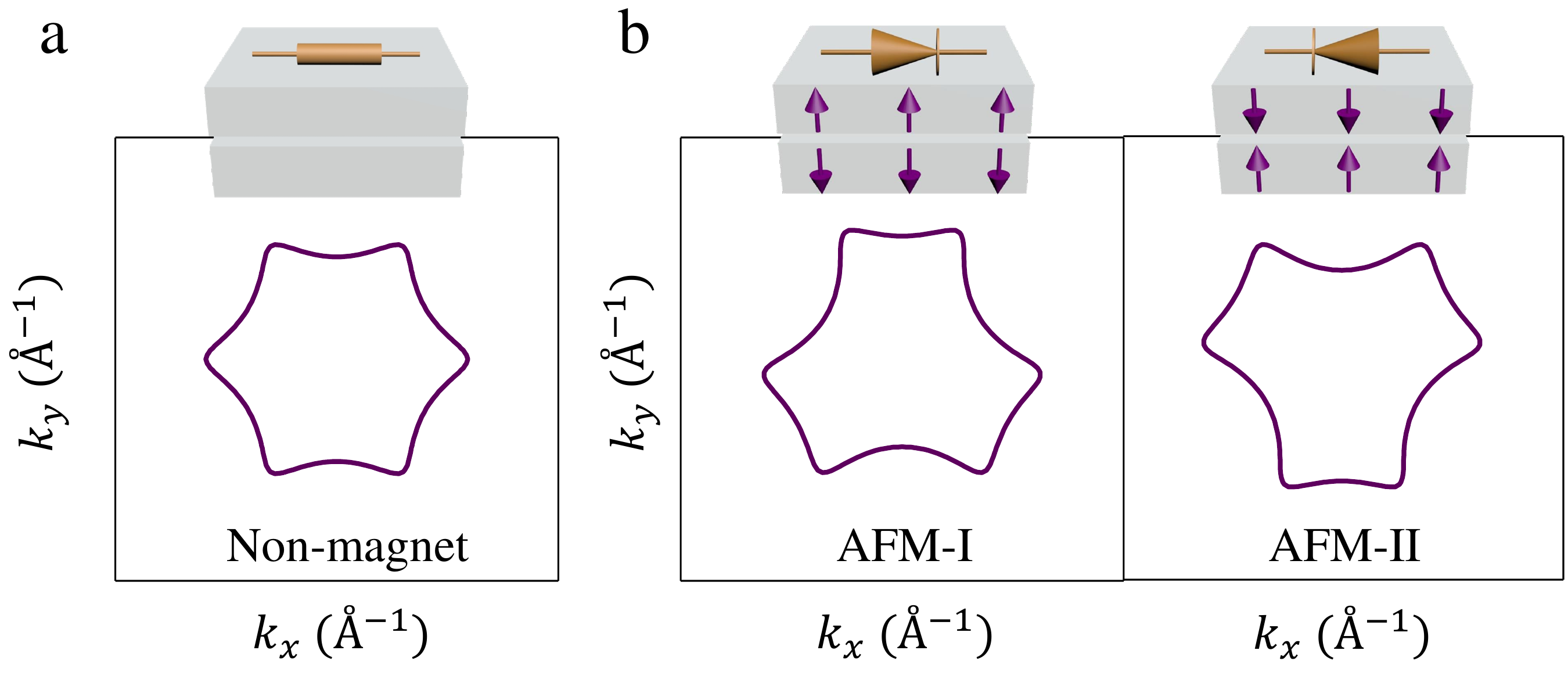}
\caption{\textbf{The asymmetric band contours induced by AFM spins in 6SL MnBi$_2$Te$_4$ at $E_f = 0.2$ eV.} \textbf{a,} The band contours of the Dirac surface states without AFM order. The band contour are symmetric with respect to $\pm k$. \textbf{b,} The band contours of the Dirac surface states for opposite AFM states. The AFM spins make the band to be $\pm k_y$ asymmetric. This AFM spin induced asymmetry is opposite for opposite AFM states. This explains the observation that the AFM diode effect signal is opposite for opposite AFM states shown in Figs.~\ref{Fig2}\textbf{e-f}, \ref{Fig4}\textbf{f} and Extended Data Fig.~\ref{hysB_Nonlinear}.
%{b}, DFT calculated band contours of 6SL MnBi$_2$Te$_4$. The band energies are color-coded from 0.05 to 0.25 eV in a 0.05 eV step. The band shows a notable asymmetry along $k_x$ direction.
}
\label{Contour}
\end{figure*}

\clearpage
\begin{figure*}[h]
\includegraphics[width=16cm]{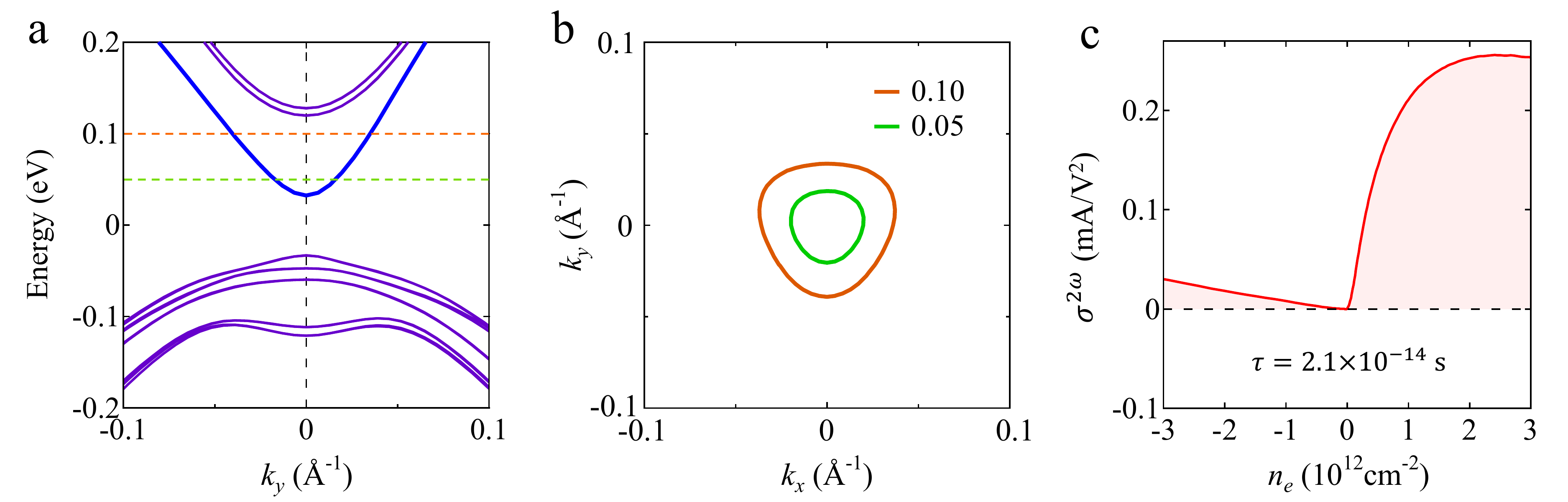}
\caption{\textbf{DFT calculated band structure of 6SL MnBi$_2$Te$_4$.} \textbf{a,} Band structure of 6SL MnBi$_2$Te$_4$ in $k_y$ direction. \textbf{b,} DFT calculated band contours at specific energies (0.05 eV and 0.10 eV), marked with dashed lines in panel (a). The low energy Dirac bands show dramatic asymmetry between $+k_y$ and $-k_y$.  \textbf{c,} DFT calculated nonlinear conductivity $\sigma^{2\omega}$ as a function of Fermi level $E_f$. We obtained the relaxation time $\tau = 2.1\times 10^{-14}$ s based on the Drude conductivity.  We note that our calculations are based on the DFT band structure of 6SL MnBi$_2$Te$_4$, which predicts a magnetic gap $\sim50$ meV. However, experimentally, the gap size is not settled \cite{Trang2020, Otrokov2019a, Hao2019gapless}. In particular, spatial inhomogeneity can smear out the magnetic gap \cite{beidenkopf2011spatial}. Moreover, the antisite defects could suppress the band gap and can increase the quantum metric effect, as predicted in Ref. \cite{kaplan2022unification}.
}
\label{Band_AFM}
\end{figure*}

\clearpage
\begin{figure*}[h]
\includegraphics[width=17cm]{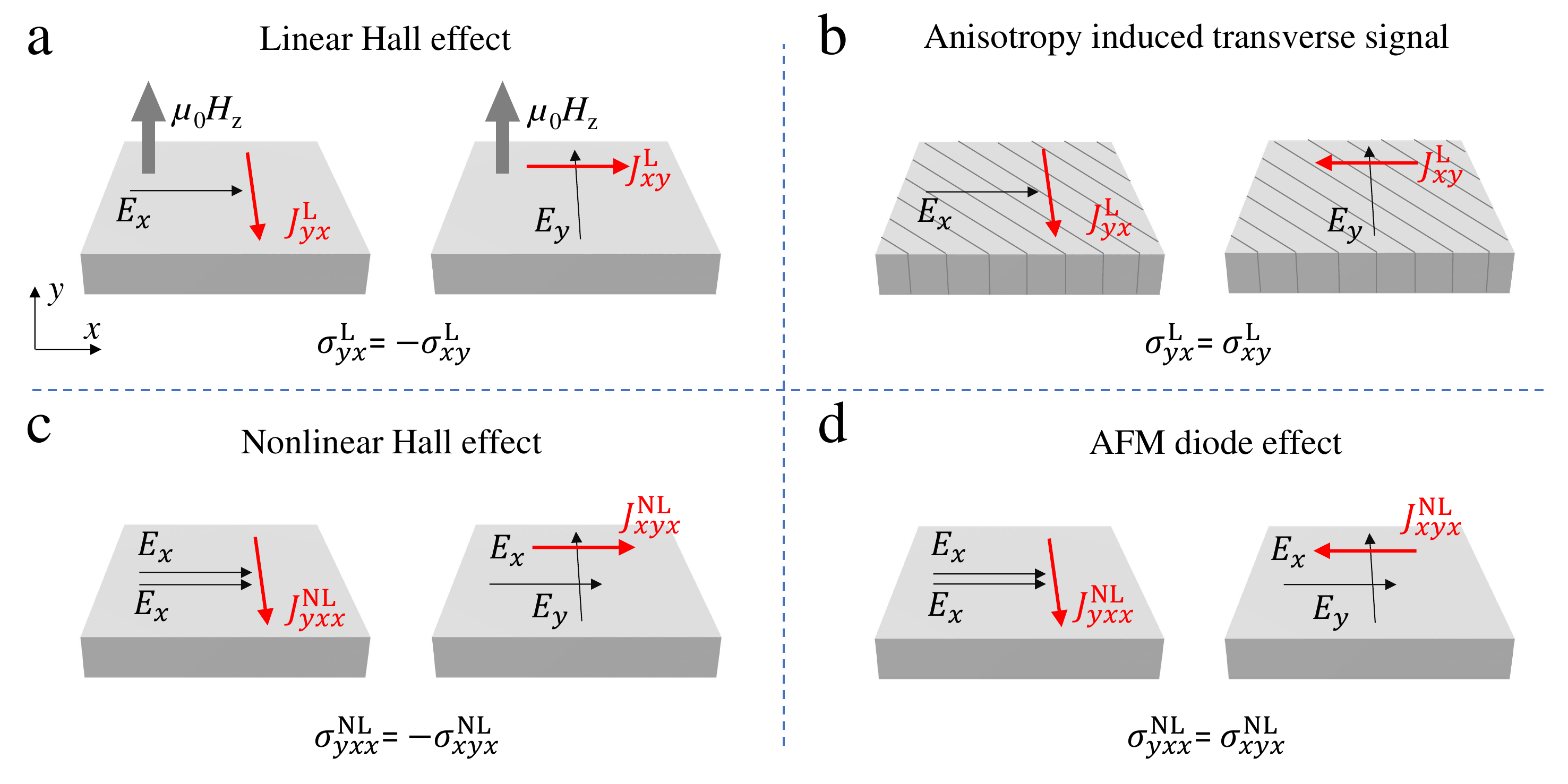}
\caption{\textbf{The exchange property of transverse conductance.} \textbf{a,b} Schematics of two different types of linear transverse conductance, the Hall effect (panel a) and the anisotropy induced transverse signal (panel b). The Hall effect occurs due to external magnetic field or internal magnetization. The anisotropy induced transverse signal occurs due to crystalline anisotropy (e.g. orthorhombic or monoclinic). These two different mechanisms can be distinguished by their exchange properties. The Hall effect is antisymmetric ($\sigma_{xy}=-\sigma_{yx}$), whereas the anisotropy induce transverse signal is symmetric ($\sigma_{xy}=\sigma_{yx}$). \textbf{c,d} Schematic of nonlinear Hall signal (panel c) and the transverse component of the AFM diode signal (panel d). The nonlinear Hall signal is antisymmetric ($\sigma_{yxx}=-\sigma_{xyx}$), while the AFM diode signal is symmetric ($\sigma_{yxx}=\sigma_{xyx}$).
}
\label{Anisotropy_Hall}
\end{figure*}

\clearpage
\begin{figure*}[h]
\includegraphics[width=13cm]{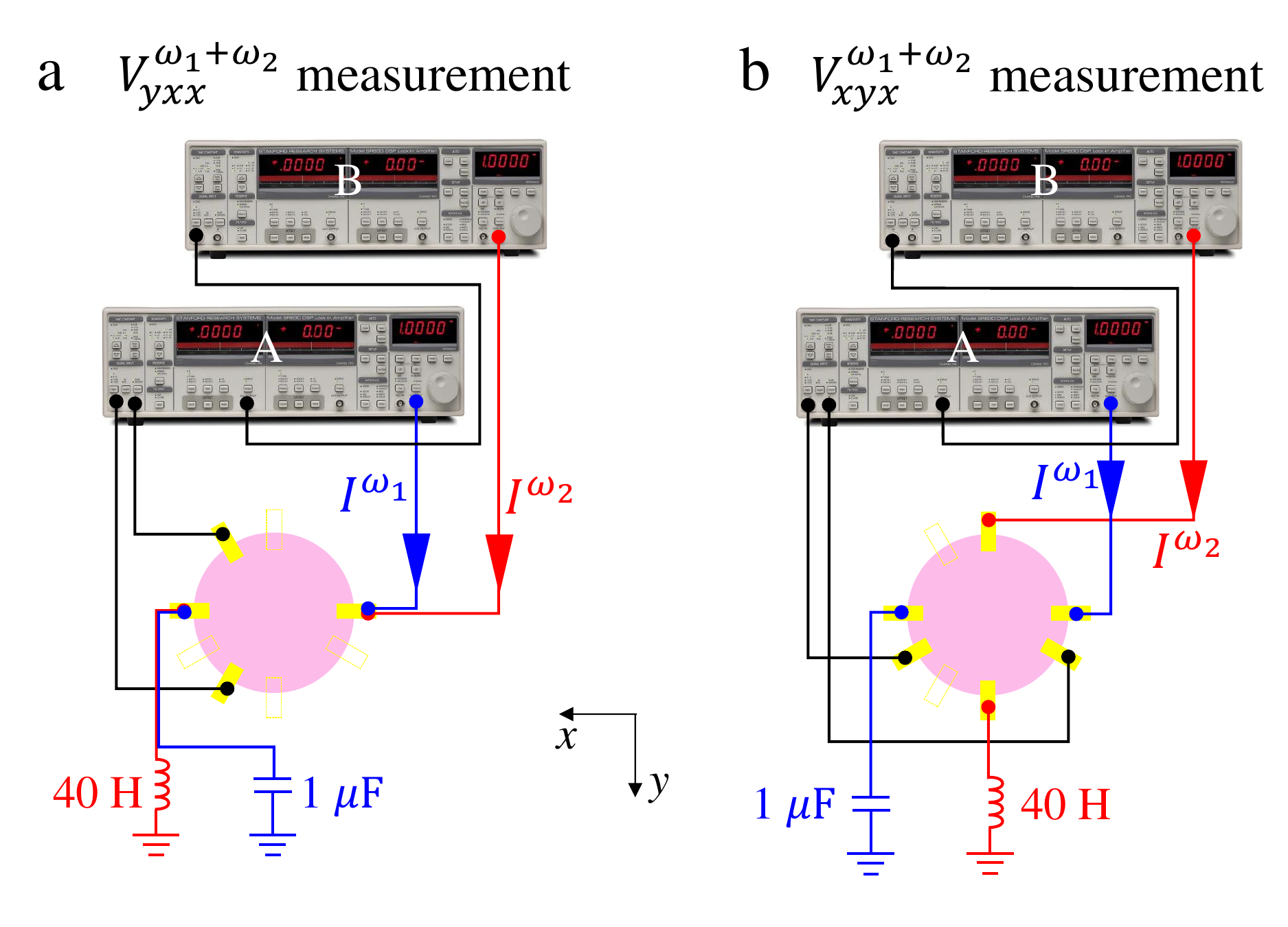}
\caption{\textbf{Electrical sum-frequency generation (SFG) measurement for 6SL MnBi$_2$Te$_4$.} \textbf{a,b,} Schematic for $V_{yxx}^{\omega_1+\omega_2}$ and $V_{xyx}^{\omega_1+\omega_2}$ measurements, respectively. $\omega_1=1.77$ Hz and $\omega_2=547.1$ Hz were used in the measurement. The capacitor and inductor were used to define the current trajectory for $I^{\omega_1}$ and $I^{\omega_2}$ (see Methods for more details).
}
\label{SFG}
\end{figure*}

\clearpage
\begin{figure*}[h]
\centering
\includegraphics[width=15cm]{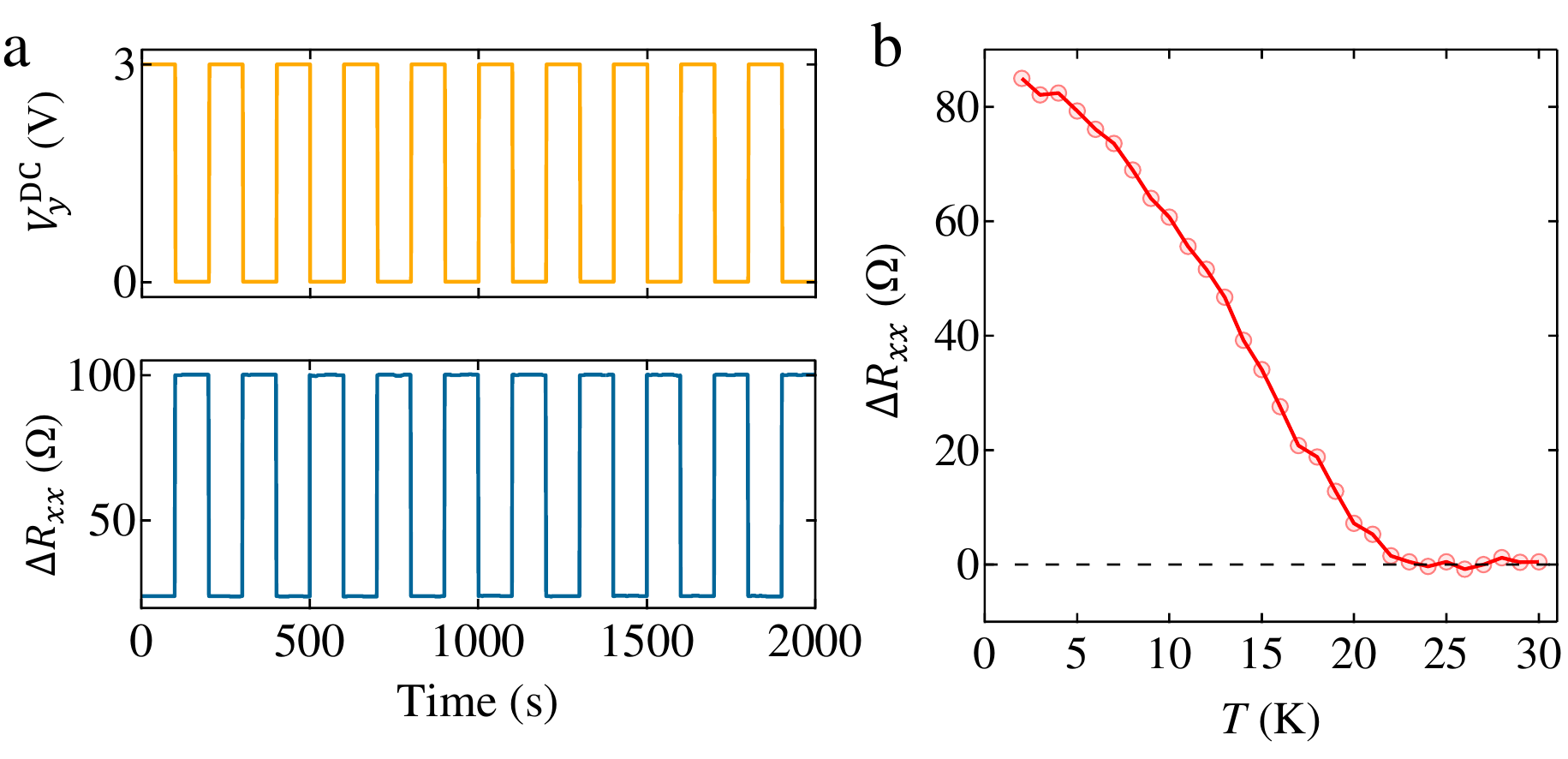}
\caption{\textbf{In-plane field effect in another 6SL MnBi$_2$Te$_4$ device.} \textbf{a}, Measured change of resistance $R_{xx}$ ($\Delta R_{xx}=V^{\omega_1}_x/I^{\omega_1}_x$) as a function of an electric field bias along $V_y^{\text{DC}}$. \textbf{b}, Temperature dependence of $\Delta R_{xx}$.
}
\label{Inplane_Field_Temp}
\end{figure*}
\vspace{1cm}

\renewcommand{\thefigure}{S\arabic{figure}}
\renewcommand{\figurename}{Fig.}
\setcounter{figure}{0}

\newpage
\vspace{3cm}
\Large
\begin{center}
Supplementary Information for\\
\vspace{1cm}
\textbf{An antiferromagnetic diode effect in even-layered MnBi$_2$Te$_4$}

\vspace{0.5cm}

\vspace{0.5cm}
%Correspondence to: suyangxu$@$fas.harvard.edu
\end{center}

\vspace{3cm}
\textbf{
This file includes:
}

\vspace{3mm}
\large

\hspace{5mm}    I. Addressing alternative mechanisms for the nonlinear signals

\hspace{10mm}	I.1. Berry curvature dipole 

\hspace{10mm}   I.2. Nonreciprocal transport in noncentrosymmetric crystals

\hspace{10mm}   I.3. Joule heating induced anomalous Nernst effect

\hspace{10mm}   I.4. Other extrinsic effects

\hspace{5mm}    II. Possible microscopic mechanisms for the nonlinear signals in antiferromagnets

\hspace{5mm}    III. Nonlinear Drude conductivity as a function of gap size

\textbf{References}

\normalsize
\clearpage
\section*{I. Addressing alternative mechanisms for the nonlinear Hall signals}

\subsection*{I.1. Berry curvature dipole}

\hspace{4mm} Both $\mathcal{PT}$ and $C_\textrm{3z}$ symmetries in even-layered MnBi$_2$Te$_4$ enforce the Berry curvature dipole to vanish. Below, we assume that both $\mathcal{PT}$ and $C_\textrm{3z}$ are weakly broken, so that Berry curvature dipole is allowed. Even so, our data below can demonstrate that the Berry curvature dipole contribution is insignificant.

\begin{figure}[h]
\centering
\includegraphics[width=16cm]{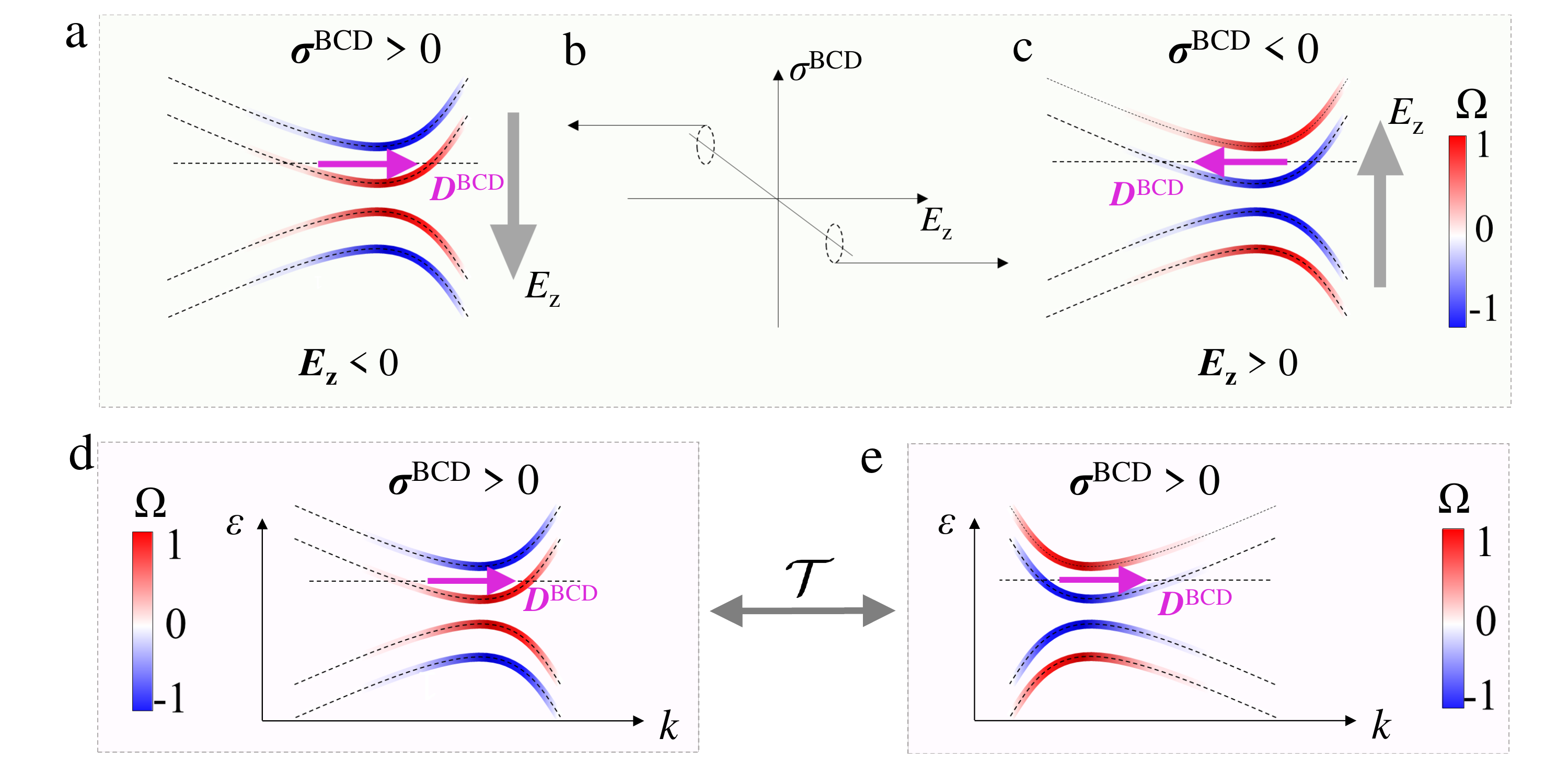}
\caption{\textbf{Berry curvature dipole and band structure of even-layered MnBi$_2$Te$_4$.}  MnBi$_2$Te$_4$ features two copies of topological Dirac surface states, top and bottom surfaces. The top and bottom Dirac surface states carry opposite Berry curvature. However, pristine even-layered MnBi$_2$Te$_4$ has $\mathcal{PT}$ symmetry. Hence the top and bottom Dirac surface states are degenerate, making the global Berry curvature and Berry curvature dipole to vanish. (\textbf{a-c}) The application of a finite electric field $E_z$ breaks $\mathcal{PT}$ symmetry. Hence the top and bottom Dirac cones become non-degenerate, leading to finite Berry curvature and Berry curvature dipole. $E_z>0$ and $E_z<0$ have opposite Berry curvature dipole. (\textbf{d,e}) Schematic illustration of the Berry curvature dipole ($D^{\textrm{BCD}}$) for the AFM-I and AFM-II, which are related by time-reversal symmetry. Because pristine even-layered  MnBi$_2$Te$_4$ has zero Berry curvature and Berry curvature dipole due to its $\mathcal{PT}$ symmetry, we assume that even-layered  MnBi$_2$Te$_4$ is subject to a vertical $E_z$, which breaks its $\mathcal{PT}$ symmetry, and hence a Berry curvature dipole is allowed.}
\label{BCD_AFM}
\end{figure}

\begin{itemize}
\item \textbf{AFM order}: The two AFM states (related by time-reversal $\mathcal{T}$) are expected to have the same Berry curvature dipole induced nonlinear Hall signal (Figs.~\ref{BCD_AFM}d,e). In contrast, our observed nonlinear signal flips sign upon reversing the AFM order (see Fig. 2e,f).

\item \textbf{Vertical $E_z$ field}: The Berry curvature dipole induced nonlinear Hall is expected to be odd under $E_z$, i.e., $\sigma^{\textrm{BCD}}(+E_z)=-\sigma^{\textrm{BCD}}(-E_z)$ (Fig.~\ref{BCD_AFM}\textbf{a-c}). By contrast, our observed nonlinear signal is symmetric around $E_z=0$ as shown in Fig. 3a.

\item \textbf{Symmetric nonlinear conductivity}: The Berry curvature dipole induced nonlinear Hall conductivity is expected to be antisymmetric upon exchanging the first two indices, i.e., $\sigma^{\textrm{BCD}}_{yxx}=-\sigma^{\textrm{BCD}}_{xyx}$. However, our data is symmetric $\sigma^{2\omega}_{yxx}=\sigma^{2\omega}_{xyx}$ (Fig. 4b), which is further consistent with the AFM group $-3'm'$.

\item \textbf{Scaling property}: For the Berry curvature dipole, the nonlinear Hall conductivity is expected to be proportional to the scattering time, $\sigma^\textrm{BCD}\propto \tau$. However, our observed nonlinear conductivity is expected to be proportional to the even order of scattering time, $\sigma^{2\omega} \propto a_0\tau^0+a_2\tau^2$.

\end{itemize}

\subsection*{I.2. Nonreciprocal transport in noncentrosymmetric polar crystals}

\hspace{4mm} Nonreciprocal transport in noncentrosymmetric polar crystals can occur with or without magnetic field. With magnetic field (or internal magnetization), nonreciprocal transport resistance can arise from $\bigtriangleup R^\textrm{NT} \propto \mathbf{I}\cdot(\mathbf{P}\times \mathbf{B})$  \cite{rectification,yasuda2016large,he2018bilinear,zhang2022non}. $\mathbf{I}$ is the applied current, $\mathbf{P}$ is the polar direction of the crystal and $\mathbf{B}$ is the  magnetic field or magnetization. To observe this nonreciprocal transport, $\mathbf{B}$ and $\mathbf{P}$ need to be perpendicular to each other. We can exclude this as follows

\begin{itemize}
\vspace{3mm}

 \item \textbf{Centrosymmetric crystal}: MnBi$_2$Te$_4$ is a centrosymmetric crystal. Therefore, there is no spontaneous electrical field polarization ($\mathbf{P} = 0$). Moreover, even-layered MnBi$_2$Te$_4$ is a fully compensated AFM ($\mathbf{M} = 0$). Therefore, in even-layered MnBi$_2$Te$_4$, the nonreciprocal transport signal is expected to be zero.

\item \textbf{Full compensated AFM}: If we assume there is a small in-plane electrical field polarization, then an out of plane magnetization is necessary to generate the nonreciprocal transport signal. However, even-layered MnBi$_2$Te$_4$ is a fully compensate AFM. The scanning NV measurements (Fig. 1e and Extended Data Fig. 1) confirm that 6SL MnBi$_2$Te$_4$ is a fully compensated AFM. 
%Moreover, the nonreciprocal transport signal is expected to be antisymmetric as a function of external magnetic field $B_z$. However, the nonlinear signal observed in even-layered MnBi$_2$Te$_4$ is symmetric as a function of $B_z$ (Fig. 3b).

\item \textbf{Vertical $E_z$ field dependence}: If we assume there is a small in-plane magnetization, then an out of plane electrical field polarization is necessary to generate nonreciprocal transport signal. However, as shown in Fig. 3a, the nonlinear signal is nonzero at $E_z = 0$ ($P \propto E_z$). Moreover, the nonreciprocal transport $\bigtriangleup R^\textrm{NT} \propto \mathbf{I}\cdot(\mathbf{P}\times \mathbf{B})$ should be odd under electric field $E_z$. However, the observed data in even-layered MnBi$_2$Te$_4$ is even under $E_z$ (Fig. 3a).

\end{itemize}

With magnetic field (or magnetization), nonlinear signal can also be induced by skew scattering due to the inherent chirality of itinerant electrons in time-reversal invariant but inversion-breaking materials \cite{isobe2020high, he2022graphene,itahashi2022giant}. Such skew scattering induced nonlinearity can be expressed as $\sigma^\textrm{skew}\approx \frac{e^3 v_\textrm{F} \tau^3 }{\hbar^2 \tilde{\tau}}$, where $v_\textrm{F}$ is the Fermi velocity, $\tau$ is the isotropic scattering time, and $\tilde{\tau}$ is the skew scattering time \cite{he2022graphene}.  One can see that $\sigma^\textrm{skew}$ scales with $\tau^3$. Below we show how our data excludes such skew scattering induced nonlinear signal.

\begin{itemize}

\vspace{3mm}

\item \textbf{$\mathcal{PT}$ symmetry}: The skew scattering induced nonlinearity is forced to be zero in a $\mathcal{PT}$ symmetric system. However, our nonlinear signal is observed in $\mathcal{PT}$ symmetric 6SL MnBi$_2$Te$_4$. The $\mathcal{PT}$ symmetry is confirmed by NV (Fig. 1e and Extended Data Fig. 1) measurements.

\item \textbf{AFM order}: The skew scatting induced nonlinear signal is expected to be even under time-reversal symmetry, i.e., $\sigma^\textrm{skew} (\textrm{AFM-I})=\sigma^\textrm{skew} (\textrm{AFM-II}) $. In contrast, our observed nonlinear signal flips sign upon reversing the AFM order (see Figs. 2e,f).

\item \textbf{Scaling property}: For skew scattering, the nonlinear conductivity is expected to be proportional to the cube of scattering time, i.e., $\sigma^\textrm{skew} \propto \tau^3$. However, our observed nonlinear conductivity is expected to be proportional to the even order of scattering time, $\sigma^{2\omega} \propto a_0\tau^0+a_2\tau^2$.

\end{itemize}

\subsection*{I.3. Joule heating induced thermoelectric effect}

\hspace{4mm} We now consider a Joule heating induced thermoelectric effect: (1) Joule heating leads to a temperature gradient $\Delta T\propto I^2R$; (2) $\Delta T$ then can lead to a current via the thermoelectric effect.

\begin{itemize}
\item \textbf{AFM order}: The typical thermoelectric effect is independent of the magnetic order. By contrast, our nonlinear signal flips sign upon reversing the AFM order. Therefore, we can rule out the typical thermoelectric effect which is independent of magnetism.
\end{itemize}

Now let us consider thermoelectric effect that depends on the magnetism, including the magneto-Seeback effect and the Nernst effect.

\begin{itemize}
\item \textbf{Full compensated AFM}: Both the magneto-Seeback effect and the Nernst effect require a nonzero magnetic field $B$ or a nonzero magnetization $M$. However, even-layered MnBi$_2$Te$_4$ has fully compensate AFM, as confirmed by our scanning NV center magnetometery measurements (Fig. 1e and Extended Data Fig. 1). Our data also show the nonlinear signal does not require external $B$ field (Fig. 3b). Therefore, we can rule out the magneto-Seeback effect and the Nernst effect.

\item \textbf{Directional dependence}: Even if one assumes that there is $B$ or $M$, the thermoelectric effect is not expected to show very strong dependence on the crystalline directions. By contrast, our nonlinear responses (Fig. 2i,j) show a clear $60^{\circ}$ periodicity. $V^{2\omega}_{\|}$ is nonzero at $\theta=60^{\circ}\times n$ but is zero at $\theta=60^{\circ}\times n +30^{\circ}$. By contrast,  $V^{2\omega}_{\|}$ is zero at $\theta=60^{\circ}\times n$ but is nonzero at $\theta=60^{\circ}\times n +30^{\circ}$. Therefore, the observed strong directional dependence, which is consistent with the AFM symmetry group $-3'm'$, demonstrates the AFM diode effect. 
\end{itemize}

\subsection*{I.4. Other extrinsic effects}

\hspace{4mm} Finally, we consider various extrinsic effects such as accidental contact junctions, flake shape, etc. 

\begin{itemize}
\item \textbf{Contact junction}: An accidental contact junction can lead to a nonlinear effect. (1) The nonlinear signals induced by contact junction should not relate to AFM states. (2) The nonlinear signals induced by contact junction should not be sensitive to N\'eel temperature. (3) The nonlinear signals induced by contact junction should not show strong directional dependence respecting the symmetry of MnBi$_2$Te$_4$.

The nonlinear signals induced by contact junction can be excluded by the following observations: (1) The nonlinear signals have opposite sign for the different AFM states (Figs. 2e,f and 4f and Extended Data Fig. 3). (2) The nonlinear signals decrease to zero when the temperature is higher than N\'eel temperature (Fig. 4g). (3) The nonlinear Hall signals can only be observed when the current applied in mirror plane $\mathcal{M}_y$ (Extended Data Fig. 5). (4)The nonlinear Hall signals show a 60$^\circ$ periodicity which is consistent with the $C_\textrm{3z}$ symmetry of MnBi$_2$Te$_4$ (Fig. 2i.j). 

\item \textbf{Flake shape}: The asymmetric global shape of the sample can lead directional movement of the carriers by colliding against the asymmetric sample boundaries, which can also induce nonlinear signals. The nonlinear signals induced by flake shape should highly depend on flake shape and should not depend on AFM states, temperature and carrier density.

The flake shape induced nonlinear signals can be excluded by following observations: The nonlinear signals are highly dependent on AFM states, temperature, carrier density and crystalline direction.
\end{itemize}

Therefore, our systematic data as a function of temperature, AFM states, crystalline direction, doping, etc. allow us to show that these extrinsic effects are not important.

\clearpage
\section*{II. Possible mechanisms for the nonlinear signals in antiferromagnets}

We briefly describe the possible microscopic mechanisms according to recent theoretical studies \cite{gao2014field, Wang2021Intrinsic,liu2021intrinsic, holder2021mixed, lahiri2022intrinsic, Smith2022momentum, ma2022anomalous, hayami2022nonlinear, kaplan2022unification,huang2023nonlinear, huang2023Scaling, atencia2023disorder}. The antisymmetric part (the nonlinear Hall effect) can arise from quantum metric ($\tau^0$) and disorder (both $\tau^2$ and to $\tau^0$). The symmetric part (the AFM diode) can arise from nonlinear Drude effect ($\tau^2$) \cite{Wang2021Intrinsic}, quantum metric ($\tau^0$) and disorder (both $\tau^2$ and to $\tau^0$).

%\vspace {2cm}
\section*{III. Nonlinear Drude conductivity as a function of gap size}

\begin{figure}[h]
\centering
\includegraphics[width=17cm]{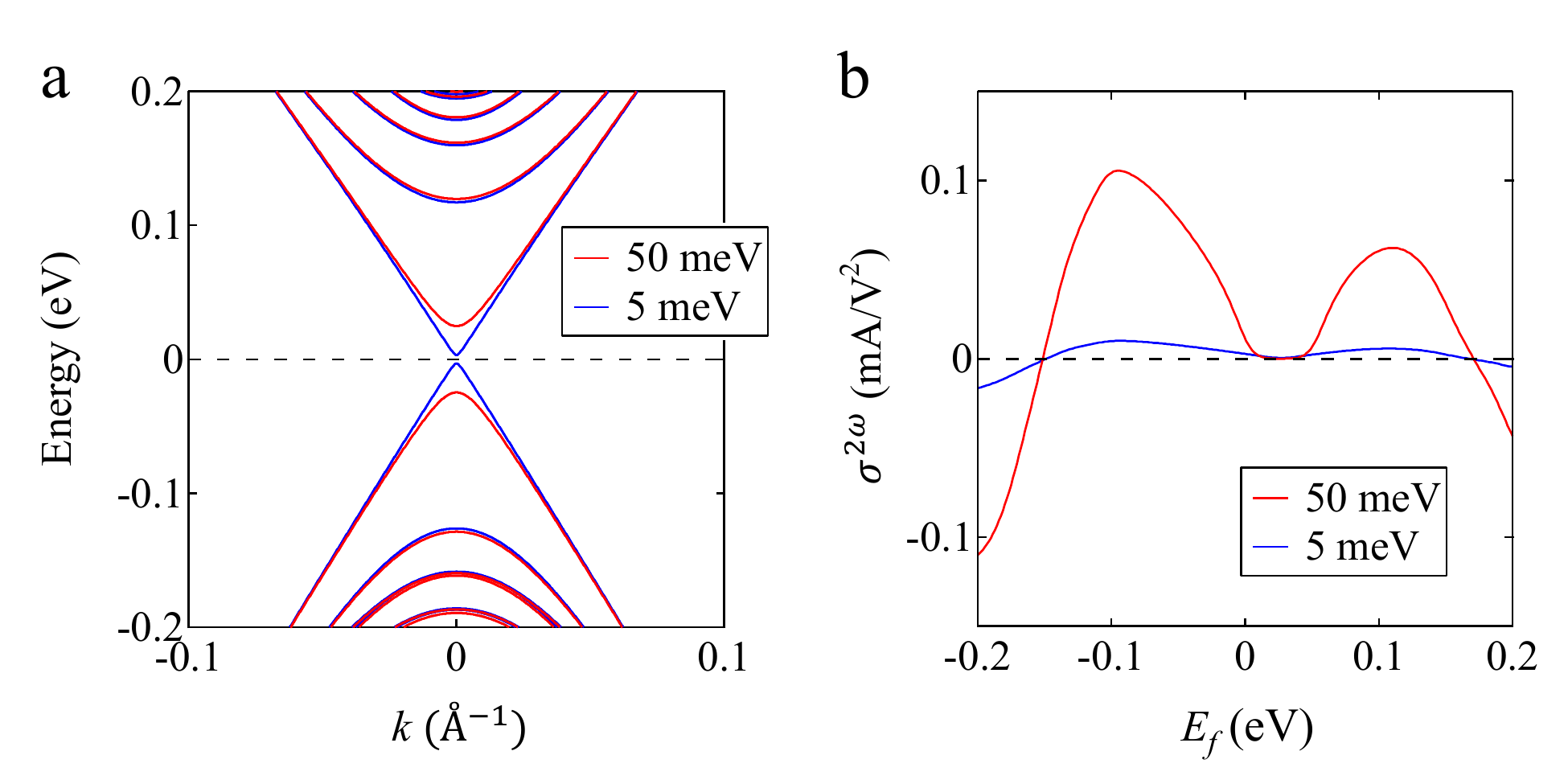}
\caption{\textbf{Magnetic gap dependence of the nonlinear Drude conductivity.}  \textbf{a,} Tight-binding band structure of 6SL MnBi$_2$Te$_4$. See details of the TB model in Refs. \cite{Zhang2020,ahn2022theory}. We tune the magnitude of the AFM order $m$ \cite{ahn2022theory}, in order to achieve two different magnetic gap values, $50$ meV and $5$ meV, with all other parameter values remaining unchanged. \textbf{b,} Calculated nonlinear Drude conductivity for the 50 meV and 5 meV gap size. We used the relaxation time $\tau = 2.1\times 10^{-14}$ s for both cases.}
\label{gap}
\end{figure}

We study how the size of the magnetic gap at the Dirac point influences the nonlinear Drude conductivity. Using the tight binding model of 6SL MnBi$_2$Te$_4$, we compute the nonlinear Drude conductivity at two different gap sizes (Fig.~\ref{gap}). By decreasing the band gap from 50 meV to 5 meV, the nonlinear Drude conductivity decreases by factor of $10$.

\normalsize
\clearpage

\clearpage

%\bibliographystyle{naturemag}
%\bibliography{Diode_03122024.bib}

\end{document}